\documentclass{article}
\usepackage[utf8]{inputenc}
\usepackage{amsmath}
\usepackage{authblk}
\usepackage{graphicx}
\DeclareUnicodeCharacter{2212}{-}

\providecommand{\keywords}[1]
{
  \small	
  \textbf{\textit{Keywords---}} #1
}

\title{Stochastic and nonstochastic descriptions of the 2019-2020 measles outbreak worldwide with an emphasis in Mexico.}
\author[1]{Luna-Banenelli, R.}
\author[2]{Vivanco-Lira, A.}
\affil[1]{Department of Medicine and Nutrition, University of Guanajuato.}
\affil[2]{Department of Medicine and Nutrition, University of Guanajuato. Department of Engineering and Exact Sciences, Open and Distance Learning University of Mexico.}
\date{Submitted on April 2020}

\begin{document}

\maketitle

\begin{abstract}
    Measles is an infectious disease caused by the Morbilivirus Measles Virus which has accompanied the human race since the 4th millennium BC, it is a disease usually concerning the paediatric population and in the past, before the advent of vaccination, almost all the population suffered from it, and in some cases the complications derived from this disease, such as central nervous involvement. Vaccination changed the course of the disease worldwide and diminished the associated comorbidities and mortality; in Mexico the vaccination program commenced in the decade of the 1970s and was successful in preventing peaks of infection. Nevertheless, due to various factors, has the world seen measles outbreaks once more, this commencing in the year 2019 and extending towards the year 2020. Here we make account of the biology and the pathophysiology of the viral infection, and present three models: one concerning the dynamics of the cases by means of a continuous method and a discrete stochastic model; one concerning the cellular compartmentalization behaviour of the virus, that is the viral tropism towards certain cell types in the host and the tendencies in extended or complicated infection; the last one concerning geographic behaviour of the virus, regarding in particular the tendencies in Mexico City, those involved at a global scale, and finally a model providing a prediction of the viral genotypes' distribution worldwide. 
\end{abstract} \hspace {10pt}

%TC:ignore
%TC:endignore

\tableofcontents

\section{Introduction}
Measles is an infectious viral disease with a Paramyxoviridae member as the aetiologic factor, the measles virus, a negative-sense single-stranded RNA virus, and it may produce dermatologic lesions as well as respiratory signs had there been no preventive measures taken, for it was by means of the introduction of the massive immunization programs that the incidence rate of this infection were reduced dramatically, nevertheless it has been seen that immunization is sometimes avoided because of misinformation or other causes, due to this fact, the rate of measles has once more risen. This virus may be transmitted by means of aerosols or in fomites in a lesser extent. It has been postulated that measles originated from modern canine distemper or rinderpest viruses in the Epipaleolithic Age (10 000 BC), and that it became a human disease in Mesopotamia in the 4th millennium BC (Retief \& Cilliers, 2010). 
\section{Biology and pathophysiology of measles virus}
Measles virus (Measles morbillivirus) is a member of the genus of viruses Morbilivirus which belongs to the subfamily Orthoparamyxovirineae, belonging itself to the family Paramyxoviridae; these viruses being negative sense RNA viruses (Wang, et al., 2018). The RNA genome of this virus is approximately 16 kbp containing six genes encoding six structural proteins: nucleocapsid (N), phosphoprotein (P), matrix (M), fusion (F), haemagglutinin (H) (the haemagglutinin protein and the fusion protein later form a heterooligomer in order to promote membrane fusion (Hashiguchi, Maenaka, \& Yanagi, 2011)) and large (L) proteins; along with two non-structural proteins: V protein and C protein (Rota, et al., 2016). It has been seen that the viral particles vary in diameter from 50 to 510 nm in its wild-type form (Liljeeros, Huiskonen, Ora, Susi, \& Butcher, 2011). Strains of these virus have been characterized in regards to the genetic variability of the H and the N proteins, 8 clades and 24 genotypes: A, B1-B3, C1-C2, D1-D11, E, F, G1-G3, H1-H2), as it was witnessed in the 2013 to 2016 measles outbreak in Cameroon, in which B3 strains featured the panorama (B3.1, B3.2, B3.3, U64582 MVi/Johannesburg.ZAF/0.88/1[D2] (Obam Mekanda, et al., 2019). Three receptor proteins have been noted in hosts for measles virus: SLAM, CD46 and nectin 4 (Hashiguchi, Maenaka, \& Yanagi, 2011). 
\subsection{MV proteins}
\subsubsection{Haemagglutinin}
This protein is a type II membrane glycoprotein displaying four domains: N-terminal cytoplasmic tail, transmembrane region, a stalk, C-terminal receptor-binding head domain; the latter exhibits a six-bladed beta-propeller fold, which is found in other paramyxoviruses and human proteins (Hashiguchi, Maenaka, \& Yanagi, 2011), the beta-propeller fold is a highly symmetrical fold with paired repeats of a four-stranded antiparallel beta-sheet motif, in the interaction between the sheets is provided the hydrophobic character of the propeller augmenting the structure’s stability (Pons, Gómez, Chinea, \& Valencia, 2003). Since the H protein lacks the neuraminidase domain, it targets SLAM, CD46 and nectin-4. This protein contains a top pocket which resembles the sialic acid binding cavity present both in haemagglutinin-neuraminidase and neuraminidase proteins, but this H protein lacks sialic acid-binding activity. The head domain homodimerizes (Hashiguchi, Maenaka, \& Yanagi, 2011), this head has been seen to further dimerize, forming then a dimer of dimers and postulations in regards to the role of the tetramer formation in promoting membrane fusion have been made (Nakashima, Shirogane, Hashiguchi, \& Yanagi, 2013). The glycosylation in this protein occurs between amino acids 168 and 238 (as seen in Edmonston type virus) but wild type viruses possess an additional glycosylation site at amino acid 416 these sites are glycosylated in an N manner. The fusion activity is triggered by the H binding to receptors, promoting conformational changes in F protein (Lin \& Richardson, 2016); another function of the H protein is to mediate virus adsorption onto the cellular membrane of the host (Saitoh, et al., 2012). Genetic variability in the H gene has been noted and appears to have a geographic pattern, with the genotypes D4 and D6  mainly found in European countries, while D3, D5, D9 and H1 in Asian countries (Saitoh, et al., 2012). This protein exhibits antigenicity and humoral immunity is directed mainly towards the H protein, being its epitopes: hemagglutinating and noose epitope (HNE), receptor-binding epitope (RBE), sugar-shielded epitope (SSE), neutralizing epitope (NE), and loop epitope (LE). The HNE is formed by amino acids from 379 to 400 and three cysteines from the noose motif forming a disulphide-constrained surface-exposed loop, this epitope is well conserved among measles virus strains. The RBE is found at the lateral side of the propeller fold, a Phe552 interacts with SLAM, and a mutation in this site induces immune escape; the epitope recognized by the monoclonal antibody BH26 (MAb-BH26) in the RBE is said to be the most important immunodominant epitope, MAb-BH26 inhibits the binding of 60\% of human serum antibodies in measles patients or vaccines. A glycosylated amino acid at the 416 position shields a specific epitope from monoclonal antibody recognition (E128 and BH99), said sugar was gained in an ancestral strain (since G3, H1 B2, B3, D6 genotypes lack a glycosylated motif in this amino acid site) (Tahara, et al., 2016).  Genetic evolution of the H protein has been witnessed, being the oldest strain: A (1948); and the most recent divergent ones: H1 and H2 (1978 and 1986, respectively); with an estimated evolution rate of $9.02x10^{-4}$ substitutions/site/year (Kimura, et al., 2015).

\subsubsection{Fusion protein.}
This protein is involved in the viral membrane fusion with that one of the host, by means of an initial conformational change induced by the H protein, decreasing the distance between the membranes and by the protein’s structural change can both membranes fuse; this process is not only important for the viral entry into the cell but as well to promote cell-to-cell spread forming syncytia (or multinucleated cells) such as the Warthin-Finkeldey cell (or reticuloendothelial giant cell) in lymphoid tissue, epithelial giant cell in epithelium, but it has been also noted that the endothelium may as well be induced to form syncytia (Takeuchi, Miyajima, Nagata, Takeda, \& Tashiro, 2003). The F protein is a type I integral transmembrane and class I viral fusion protein which homotrimerizes, each F monomer is initially synthesizes as a long precursor F0, then by means of the Golgi network, this precursor is cleaved into F1 and F2, then linked by means of two disulphide-bonds from F1 to F2 subunits, this process is mediated by furin (Plattet, Alves, Herren, \& Aguilar, 2016) this protein is another type I transmembrane protein which exhibits homology with other proprotein convertases, belonging to the subtilisin superfamily of serine endoproteases, this protein shows conserved residues in the catalytic triad, this protein is pH-dependent with a range from 5 to 8, this dependent on the substrate to be cleaved (Thomas, 2002) it may be inhibited by decanoic acid and pirfenidone (TyersLab, 2020). Trimerization of the F protein occurs at the endoplasmic reticulum, in the same place where H as well oligomerizes, promoting the formation of membrane fusion machineries and F-H heterooligomers, however this interaction’s affinity diminishes once F0 is cleaved. There are well-conserved regions: cytoplasmic and transmembrane domains, two heptad repeat regions (HRA, HRB, HRC), N-terminal hydrophobic segment. Since this protein is a class I viral fusion protein, it is initially folded in an intrinsically high energy structure (prefusion state) maintained by means of a high energy barrier carrying then on spontaneous (from a low entropy state) changes in its structure unto a stable low energy, high entropy state (postfusion state) with a central core six-helix bundle domain. The F protein (in other paramyxoviruses) unveils a short stalk region with three HRB regions supporting a globular domain, this latter globular domain with three subdomains (DI, DII, and DIII), this DIII domain suffers drastic changes when transiting from pre- towards postfusion states while DI and DII suffer less dramatic conformation movements. The conformational changes induced in the F protein, by means of the anchoring of the transmembrane domain in the viral envelope and the closure of the fusion peptide with the host membrane, ensure changes in the membrane curvature and an exergonic membrane fusion with the concomitant pore formation. Several F protein inhibitors have been proposed, such as peptides, antibodies and small molecule; in this small molecule class, we have: AS48 with a molecular mass of $299.09 Da$, and a polar surface area of $1.18 nm^2$ (Royal Society of Chemistry, 2020)), JNJ-2408068 with a molecular mass of $394.248108 Da$, and a polar surface of $0.92 nm^2$ (Royal Society of Chemistry, 2020)), JNJ-49153390  with a molecular mass of $504.4 Da$ and a polar surface area of $0.97 nm^2$ (Royal Society of Chemistry, 2020)), TMC-353121 , BTA9881, BMS-433771 (Plattet, Alves, Herren, \& Aguilar, 2016). 
\subsubsection{Phosphoprotein.}
This protein attaches to the ribonucleoprotein complex of measles virus (that is, RNA bound to the nucleocapsid protein); it is as well a cofactor for the L protein. This phosphoprotein (P) has two domains: an N-terminal domain (playing a role in viral replication) and a C-terminal domain (involved in both replication and transcription). The N-terminal domain associates with the newly synthesized nucleoprotein $N^0$ functioning as its chaperone. Fully functional N provides two interactions with the N-terminal domain of the P protein, say the N core domain with the N-terminal domain and the N tail domain with the C-terminal domain interact, mediating the binding of the P protein to the newly synthesized N protein, this prevents the binding of the N protein with host RNA. The XD region of the P protein induces proper folding of the N tail domain. The P protein interferes with the JAK1/STAT pathway, degrading STAT1 or preventing the phosphorylation of JAK1 or STAT1; JAK1 and STAT1 both of them activate IFNA, IFNB, IFNG and IL10, as well as several members of the gp130 family of proteins (IL6, IL11, CNTF, CT1, GCSF, LIF, OSM) (Rawlings, Rosler, \& Harrison, 2004); due to this fact could we approach the lack of immune response towards the virus by means of JAK/STAT signalling pathway activators or to directly administrate substrates of the downstream signalling pathway; moreover, STAT1 has shown to promote epigenetic changes by means of the H3C1 histone, therefore, could Measles virus be able to change the epigenetic landscape of the host (Rui, et al., 2016). This XD region has been modelled to possess 53 amino acid residues, 64 water molecules, and (in the crystallization method) one molecule of 2-n-(cyclohexylamino)ethanesulfonic acid (Johansson, et al., 2003), due to this binding could we look into a similar molecule which could show promise in P protein inhibition such as 4-methyl-5-(beta-hydroxyethyl)thiazole monophosphate which shows thiamine-phosphate diphosphorylase activity (PubChem, 2020). The P protein may also display anti-apoptotic features (for it has been seen that an increase in both BCL2 and BCLXL associates with an increase of the P protein concentration; as well as an increase in vimentin expression by means of the P protein (Bhattacharjee, Kumar Jaiswal, \& Kumar Yadava, 2019). 
\subsubsection{Nucleoprotein.}
This N protein is composed of two regions, one spanning the amino acids from 1 to 391, the N core region; and another one spanning the amino acids from 392 to 525, the N tail region; the N core region is an ordered one, while the N tail region is an intrinsically disordered one (these intrinsically disordered regions of domains are found in nearly all species and function to allow the same protein to interact with different substrates and thus, to be part of different signalling networks (Wright \& Dyson, 2015)). The N core region contains two domains (NTD and CTD) which are flanked by N- and C-terminal arms; N oligomerization is mediated by means of exchange of the NT arm and the CT arm; the RNA binding site is on the groove between two N core domains. This protein composed as well by a tetramerization domain, MD, forming: four-helix coiled-coil, C-terminal domain (extreme), XD, disordered regions; the L protein (RNA polymerase) complexes with P and then attaches to NC by means of the XD domain in P and the molecular recognition element. P plays an important role in N metabolism, for it acts as a chaperone to keep the newly synthesized N protein in an RNA-free monomeric form, as well as it positions the polymerase complex for polymerization (Guryanov, Liljeeros, Kasaragod, Kajander, \& Butcher, 2016). The association of N core with RNA forms helices, each helix is composed of 12.34 nucleoprotein subunits per turn, with a slope of 4.9 nm and an outer diameter of 19 nm; the RNA winds around the N protein in a bobbin-like fashion (Gutsche, et al., 2015), this binding of the RNA to the N protein arises from the phosphates and 2’ oxygen of the ribose, but not interactions dependent on sequences, in regards to the kinetics of this binding, each N protein protomer binds to 6 RNA bases (Guseva, Milles, Blackledge, \& Ruigrok, 2019).  
\subsubsection{V protein.}
This protein is encoded by the P gene along with two other proteins (say R and C), containing a unique Cys-rich C-terminus; C, V and R are identical for the first 231 amino acids, afterwards there exists a divergence induced by various mechanisms. The cysteines at the C-terminus may mediate zinc binding (these type of proteins (zinc-binding proteins) require zinc to become stable and the concentrations required of zinc for such stability is $[50,100] \mu M$ (Terada \& Yokoyama, 2015)), being able then to contain zinc-finger-like domains (the first zinc-finger protein was found to be able to DNA by means of the cysteine repeats within the sequence, with some of these proteins displaying transcriptional or ubiquitin activity (Cassandri, et al., 2017)). Two functions of this protein have been widely recognized: induce changes in the STAT pathway (Parks, et al., 2006) this blockage has been seen to be by means of interaction with STAT1, STAT2, IRF9 and STAT3, blocking of STAT2 phosphorylation and preventing activation of JAK1 (Pfaller \& Conzelmann, 2008), interfering with interferon synthesis and inhibiting viral RNA synthesis, it has also been noted to be able to bind to RNA possibly by means of the zinc-finger-like structure it possesses; the role of this protein in RNA synthesis inhibition is possibly to maintain a correspondence of viral protein concentration in the cytoplasm to RNA synthesis or to maintain a ratio of genome to antigenome (Parks, et al., 2006). The V protein has shown high affinity to both IKKA and IRF7, it also competes with IRF7 in phosphorylation; preventing interferon induction and promoting to the immunosuppressive seen in measles patients (Pfaller \& Conzelmann, 2008). 
\subsubsection{L protein.}
This protein has properties of nucleotide polymerization, 5’ capping, and polyadenylation (Ito, et al., 2013) in this case the polyadenylation serves as a gene end signal, proceeding over three nucleotides without transcribing them and restarting the transcription when recognizing the gene start signal (Bloyet, et al., 2016). As we have previously commented the V protein inhibits RNA synthesis but also the C protein does it so (which also derives from the P gene) (Parks, et al., 2006); the replication machinery of measles virus is comprised by the N, P and L proteins (Bloyet, et al., 2016). The L protein derives from a 6644 nucleotide gene, translated into a 2183 amino acid protein (with a molecular weight of around 250 kDa) with two highly variable regions (hinges): H1 (from amino acid 607 to 650) and H2 (from amino acid 1695 to 1717) which may form protein boundaries. In this protein there lay six conserved sequences, being H1 a region of low homology. Furthermore, it has been seen that by tagging H1 with fractions of MYC may compromise the whole L protein polymerization function, nevertheless this occurs not when the tag was added to the H2 region; posing the idea that the highly variable region is a quite important factor for proper protein function, this possibly by means of a reorientation of the critical catalytical domain (D2), domain which itself shows a high degree of conservation (78.9\%) (Paul Duprex, Collins, \& Kima, 2002). The L protein might interact with SHCBP1 as not enough information was available to confirm or deny such interaction, the latter protein being inhibited by the C protein in order to decrease interferon response (Ito, et al., 2013), SHCBP1 is a protein linked to cell proliferation by means of regulation of cell cycle progression and apoptosis due to SHCBP1 interacting with CDK4-cyclin D1 cascade, and suppression of caspase-3, and caspase-dependent apoptotic pathways (Dong, Yuan, Yu, Tian, \& Li, 2019); promoting then an invasive phenotype; thus through the C protein (and likely, the L protein) inhibition of SHCBP1 is measles virus available to promote further cell death or cell cycle progression arrest in invaded cells. 
\subsection{Host proteins.}
\subsubsection{SLAM.}
These SLAM (signalling lymphocytic activation molecule) receptors are a family of cell surface type I transmembrane glycoprotein receptors on exception of SLAMF2 (with a glycosylphosphatidylinositol membrane anchor) (Detre, Keszei, Romero, Tsokos, \& Terhorst, 2010), also a subfamily of the CD2-like family of proteins (Yurchenko, et al., 2018). These proteins contain an N-terminal V-Ig domain and a C-terminal C2-Ig domain, nevertheless, SLAMF3 is made up only by two V-Ig/C2-Ig domains. This family of proteins is comprised by 9 different receptors, these proteins are their own ligands and bind in a homophilic manner, with SLAMF1 as the only member of the family of proteins which is known to interact with measles virus (Detre, Keszei, Romero, Tsokos, \& Terhorst, 2010). SLAMF1’s gene is found at the chromosomal location 1q23.3 and exhibits 8 exons, it has been seen to be expressed in the following tissues: lymph node, spleen, appendix, bone marrow, colon, duodenum, gall bladder, stomach, urinary bladder and others; this protein is associated with treatment refractory schizophrenia and to susceptibility in inflammatory bowel disease (NIH, 2020). This family of proteins’ importance is seen when the phenotype of SHD21A -/- is witnessed, provided the fact that SH2D1A regulates SLAM members signalling, inducing X-linked lymphoproliferative disease , in which there is an innate immune response deficiency towards EBV (Huang, et al., 2016), promoting fatal mononucleosis, fatal haemophagocytic lymphohistiocytosis (due to EBV, lymphomas, antibody deficiency, other causes of immune dysregulation) (Filipovich, Zhang, Snow, \& Marsh, 2010). SLAMF1 may mediate signalling via the TNF family and antigen receptors and regulate innate immune responses, as it is seen in SLAMF1 -/- bone marrow-derived macrophages which produce less ROS in \textit{Escherichia coli} response, since SLAMF1 promotes NOX2 activity by complexing with beclin-1, PIK3C3 (VPS34), VPS15, and UVRAG; and that by means of its interaction with TLR4 it induces TRAM-TRIF-dependent signalling (Yurchenko, et al., 2018). In T cells, SLAMF1 is expressed on immature CD4(-) CD8(-) thymocytes but high titres of expression are shown in CD4(+) CD8(+) thymocytes, and not present at the surface of monocytes, but seen on their activated form (monocytes/macrophages lineages), nevertheless, in non-activated monocytes, SLAMF1 is seen expressed in their cytoplasm in the recycling compartments (this subcellular region is localized near the microtubule-organizing centre at the perinuclear region (Xie, et al., 2015)); also seen expressed in mature dendritic cells; neutrophils, eosinophils and erythrocytes are SLAMF1(-); in B cells, the SLAMF1 expression is seen promoted in the stage towards plasma cells differentiation; also seen in human-induced pluripotent stem cells; in regards to malignancies, it has been seen expressed in malignant T cells in cutaneous T-cell lymphoma and Sèzary syndrome (Gordiienko, Shlapatska, Kovalevska, \& Sidorenko, 2019). In a cellular model of chronic lymphocytic leukaemia, a deficiency in SLAMF1 would induce increased expression of CXCR4, CD38 and CD44, affecting chemotaxis by CXCL12; SLAMF1 also appears to have a role in autophagy due to the induction of phosphorylation of p38, JNK1/2, and BCL2 (promoting autophagy) and SLAMF1-deficient cells appear to display resistance to autophagy-inducer drugs (Bologna, et al., 2016); it is likely that by means of this mechanism there is a display in lymphoma regression in mice (Chung, 2001). SLAMF1 binds to H in a $1:1$ ratio with a $k_d=0.52\pm0.21\ \mu M$ and exhibits a $10.5 nm^2$ contact area with four components of the binding interface: salt bridges of Asp505 and Asp507 to the acidic patch in H at Lys77 and Arg90; Arg553 at H to Asp350 at H and Glu123 at SLAMF1 and two hydrophobic interactions of H Phe55s and Pro554 with His61 and Val63 at SLAMF1; beta-sheet between Pro191-Arg195 at H and Ser127-Phe131 at SLAMF1; the 4th site is hydrophobic; this interaction diminishes the distance between the host and viral membranes to around $14 nm$, followed by a conformation change in F which allows F to interact with the target cell membrane and ending with membrane fusion (Hashiguchi, et al., 2011). 
\subsubsection{CD46.}
The gene encoding for this protein is found at chromosomal location 1q32.2 with 14 exons; this protein is a type I transmembrane protein and part of the complement system by inactivating C3b and C4b, thereby protecting the cell from damage by the complement system; it also acts as one of the receptors for measles virus, human herpesvirus-6 (NIH, 2020), some adenoviruses, \textit{Streptococcus pyogenes}, \textit{Neisseria gonorrhoeae} (Yamamoto, Fara, Dasgupta, \& Kemper, 2013), \textit{Neisseria meningitidis}, \textit{Escherichia coli}, \textit{Fusobacterium nucleatum} (Liszewski \& Atkinson, 2015); it is as well associated with the fusion of spermatozoa with oocytes and CD46’s gene mutations are associated with susceptibility to haemolytic uremic syndrome (NIH, 2020). The location of this gene is in the gene cluster named ‘regulators of complement activity’ on chromosome 1 (Yamamoto, Fara, Dasgupta, \& Kemper, 2013), said cluster spans $21.45\ \ cM$ of this chromosome with more than 60 genes which share an ancestral motif; they are composed of a repeating unit, the complement control protein module (comprising a sushi domain) however CD46 lacks the DAA fragment, through which other complement-related proteins enhance the spontaneous decay of the convertases (Liszewski \& Atkinson, 2015). This protein is expressed in four isoforms from alternative splicing mechanisms, and the four of these isoforms show C3b and C4b binding sites, following this complement-proteins binding domains we find a glycosylated region (in which alternative splicing may occur), then a region of unknown function, a transmembrane anchor, and two cytoplasmic domains: CYT1 and CYT2. The resulting isoforms of this protein are: C1, BC1, C2, BC2. When deficiencies in the CD46 gene exist, an overactivation of the complement system may occur, promoting C3b deposition on host endothelium and systemic microthrombi formation (such as atypical haemolytic uremic syndrome and age-related macular degeneration) (Yamamoto, Fara, Dasgupta, \& Kemper, 2013); genetic variations of CD46 have been associated with various inflammatory disorders and not only with the haemolytic uremic syndrome but furthermore with systemic sclerosis, earlier development of nephritis, HELLP syndrome, recurrent miscarriage, C3 glomerulonephritis, common variable immunodeficiency, preeclampsia, systemic lupus erythematosus (Liszewski \& Atkinson, 2015). The binding of measles virus to CD46 may not only promote the fusion with the membrane of the host but as well to facilitate the breach of immune barriers of the host or (and) to create a favourable immune microenvironment. In epithelial cells, CD46 is phosphorylated by YES1 (homolog of the Yamaguchi sarcoma virus oncogene) tyrosine kinase member of the Src family and may promote cytoskeletal rearrangements that favour infection by some pathogens by its interactions with ezrin; when CD46 interacts with DLG4 (membrane-associated guanylate cyclase (MAGUK)) the actin cytoskeleton suffers changes and the epithelial cell loses its polarity, the changes in the cytoskeleton are driven by autophagosomes; CD46 when interacting with DLG4 may also come into play with SCRIB (scribble planar cell polarity protein) playing then a role in mitotic spindle formation and epithelial tumour formation. In monocytes it has been seen that upon CD46 activation by the H protein, there exists a down-regulation of IL12 (Yamamoto, Fara, Dasgupta, \& Kemper, 2013) which displays structural similarities to the IL6 family (as well as with IL31 and IL23) and uses a gp130 ortholog to transduce its signal by means of the JAK/STAT pathway, and in IL12 case, uses JAK2 and STAT1, STAT3, STAT4, and STAT5 proteins in the transduction of its signal, IL12 production is intrinsically bound to dendritic cells’ ability to recognize pathogens and promote Th1 responses, as well as to mediate resistance to intracellular infections, for instance, IL12 -/- mice were seen to be more susceptible to pathogens such as: \textit{Leishmania}, \textit{Plasmodium}, \textit{Toxoplasma}, \textit{Cryptococcus}, \textit{Francisella} and \textit{Mycobacteria} (Tait Wojno, Hunter, \& Stumhofer, 2019). The affinity of CD46 with the H protein was computed to be $k_d=0.2\ \mu M$, the r.m.s. corresponding to the superposition of the dimer has been noted to be equal to $0.133 \ \ nm$ while the interacting surface areas equal in average $10.775\ \ nm^2$ (Santiago, Celma, Stehle, \& Casasnovas, 2009). 
\subsubsection{Nectin-4}
The gene encoding this protein is found at the chromosomal location 1q23.3 with 9 exons within the genetic structure, member of the nectin family, from this we remember that the cell-adhesion molecules belong mostly to 4 classes of molecules: integrins, selectins, immunoglobulin superfamily, and cadherins; the nectin family of proteins is found within the immunoglobulin superfamily, this family is featured by a single-pass type-I membrane glycoproteins; in nectin-4 and other splice variants, the afadin-binding motif is not conserved, being this motif the one responsible for binding with the PDZ domain of afadin (afadin is a protein involved in signalling and organization of cell junctions (NIH, 2020)). It has been seen that nectin-4 displays heterophilic interactions with nectin-1 and homophilic interactions with itself, this plays a role in cell-cell adhesion where one nectin molecule protrudes from one cell and encounters another nectin molecule protruding from a different cell, associating thereof. Of special remark, nectin-4 is not the only member of the nectin family which shows affinity for viral proteins, for both nectin-1 and nectin-2 have exhibited interaction with proteins of herpes simplex viruses (HSV1 and HSV2) (Samanta \& Almo, 2015). Within the protein we may find two Ig-like C2-type domains and one Ig-like V-type domain; involved in cell adhesion being a single-pass type I membrane protein. It possesses a soluble form through proteolytic cleavage by means of the metalloproteinase ADAM17/TACE. Associations have been made of this gene’s base pair substitutions with ectodermal dysplasia-syndactyly syndrome type 1 (NIH, 2020). Furthermore, nectin-4 has been used as a tumour marker and has been positively correlated with both cell proliferation and angiogenesis, with an increase in nectin-4 expression meaning a concomitant increase in VEGF expression (Nishiwada, et al., 2015). Nectin-4 has been seen to effectively bind to an anti-nectin-4 antibody, say enfortumab vedotin, which may show potential in diminishing measles virus infectivity by inducing conformational changes in nectin-4 and thereof, decreasing the amount of possible receptors to use in the host (Challita-Eid, et al., 2016). 
\subsection {Insights into pathophysiology.}
After viral entry into host cells, the viral RNA is sensed intracellularly by DDX58 and MDA5 (Rota, et al., 2016) where DDX58 is a DExD/H-box helicase, a DEAD box protein, the ones have RNA helicases, implicated in RNA binding and alterations of RNA secondary structure; DDX58 contains a caspase recruitment domain (CARD), involved in the recognition of viral double stranded RNA (NIH, 2020); DDX58 has exhibited TLR4-dependent activation by means of LPS, therefore playing a role in phagocytosis, for it has been seen that DDX58 depletion leads to macrophage phagocytosis deficiencies (Kong, et al., 2009), this TLR4-stimulated DDX58 pathway furthermore promotes TNFA expression; DDX58 also induces IFN type I expression by means of a TLR3-independent pathway, and by means of its CARD and the ATP-dependent catalytic centre domains of DDX58 which activate IRF3, NFKB and IFNB (Wang, et al., 2008) this NFKB and IRF3 activation is possible by means of DDX58 interaction with mitochondrial protein MAVS (Zeng, et al., 2010), a mitochondrial antiviral signalling protein which regulates expression of beta interferon and contributes to antiviral immunity (NIH, 2020), MAVS regulates the recruitment of molecules involved in mitochondria-associated complexes: the adaptors TRAF3, TANK and TRADD; the kinases TBK1, IKKE; which in turn promote the transcription of IRF3 and IRF7 with a subsequent synthesis of IFN type I; whereas NFKB augments the production of IL1B and IL6. DDX58 has shown to activate inflammasomes in a similar manner in which double-stranded DNA does so with AIM2 inflammasomes (with double-stranded DNA viruses), for when DDX58 associates with ASC induces caspase 1 activation (Poeck, et al., 2010), therefore being part of the NLRP1B inflammasome and promoting pyroptosis and IL1B expression by means of proteolytic cleavage of both gasdermin D and Pro-IL1B (Broz \& Dixit, 2016). MDA5 or IFIH1 (interferon induced with helicase C domain 1) is a DEAD box protein with a putative RNA helicase, involved in RNA secondary structure changes, it is upregulated in response to IFNB and mezerein (PKC-activating compound) (NIH, 2020) also interacts with MAVS in a similar manner in which DDX58 does and induces NLRP1B inflammasome activation with a promotion in caspase 1 activation and finally both pyroptosis and IL1B expression (Dias Junior, Sampaio, \& Rehwinkel, 2018). It has been seen that Laboratory of Genetics and Physiology 2 (LGP2) is able to respond to viral DNA when the measles virus vaccine is administered (by means of recognition of 5’ copy-back defective interfering genomes) (Mura, et al., 2017). 
Measles evades immune response by means of its V, C and P proteins; where the V protein binds to MDA5 and LGP2, inhibiting IFN synthesis (Rota, et al., 2016); LGP2 is a DExH-box helicase 58 (NIH, 2020) but in distinction of both MDA5 and DDX58, LGP2 lacks a CARD domain; LGP2 regulates the activity of both MDA5 and DDX58 in response to viral RNA; LGP2 shows a greater affinity for double-stranded RNA than DDX58 and may prevent the latter protein from binding to viral RNA and could even inhibit DDX58 function; this regulation of functions of DEAD box proteins is mediated by LGP2 and its partners, some of which are well-known, such as: DICER1, PKR, NKRF, STAU1, DHX30, XRN2; some others have been recently noted: MOV10, SRRT, RBM4 and PACT; these partners and the master regulator LGP2 could promote RNA silencing as an antiviral response, promote MDA5 function, diminish DDX58 activity and thus could play a role in decreasing pyroptosis of infected cells by means of NLRP1B inflammasome formation inhibition (Sanchez David, et al., 2019). Moreover, has it been seen that LGP2 increases beta-catenin levels and decreases GSK3B expression (Zhou, et al., 2019) which would phenotypically promote cell proliferation (Gopalkrishna Pai, et al., 2017); therefore should LGP2 be inhibited by means of measles virus’ proteins, there would be a dysregulation of DDX58 functioning (promoting NLRP1B inflammasome and pyroptotic death) and a decrease in cell viability due to an increase in GSK3B and decrease in beta-catenin expression; corresponding with the lymphopoenia seen in patients with measles infection, with decreased numbers of: CD4(+) T cells, CD8(+) T cells, B cells, neutrophils, monocytes (Okada, et al., 1999) due to this fact does there exist a tendency towards opportunistic infections such as those seen with other CD4(+) T cell deficiency: human herpes viruses (Orren, et al., 1981), CMV, mycobacteria, toxoplasma, salmonella (Ul Haq Lodhi, Imam, Umer, \& Zafar, 2017), pneumocystis (seen in patients with concomitant HIV and measles infection (Rafat, et al., 2013)), candida, among others. A myeloid suppression has also been seen, and may be a reason for which adolescents and adults tend to show more complications derived from measles infections than children or newborns, due to the rate of proliferation and differentiation from bone marrow (Manchester, Smith, Eto, Perkin, \& Torbett, 2002). In an interesting fashion, individuals with a susceptible genetic background and environmental features could promote the induction of a cytokine storm	(Tisoncik, et al., 2012) should they suffer from a measles infection, by means of a lack of control of NLRP1B inflammasome, IL1B and TNFA production, inhibition of LGP2 and others, in whom we may observe hypotension (Hahné, et al., 2016) and pulmonary fibrosis (Montella, Santamaria, Maglione, \& Ciofi degli Atti, 2009). As we have previously discussed, the tropism of the virus is determined by the affinity of the H protein towards three host receptors, say CD46, nectin-4 and SLAMF1; the expression of either of these in the cellular surface could promote inhibition of cellular proliferation and promotions in apoptotic pathways, the cellular tropism or compartmentalization of the virus is further discussed in the ‘Models’ section in which we may realize the long-term clinical consequences of these tropism. 
\section{Clinical features. }
Infection by MV has been seen to begin with an incubation period in which the virus replicates within myeloid and lymphoid cells, thus establishing a systemic infection (Rota, et al., 2016). Measles is an acute febrile disease. Symptomatically it begins at the prodromal phase of the infection, lasting from 2 to 4 days, in which fever (greater than or equal to 311.8 K), cough, coryza, conjunctivitis are seen, in a rather similar way to a higher airways infection. In the 2 to 4 days following the onset of the fever the rash appears, an erythematous maculopapular exanthema, starting at the head and neck to continue towards the thorax and limbs. During the following 3 to 5 days, the rash in different parts of the body wanes in the order in which it appeared, and the full recovery starts within the next seven days in non-complicated cases  (Moss, 2017) (Strebel \& Orenstein, 2019). Koplik stains, small white to blueish white plaques, appear at the oral mucosa and are seen in 70\% of the cases, they may appear from 1 to 2 days before the rash, and may be from 1 to 2 days after the rash has disappeared, they are considered pathognomonic (Rota, et al., 2016) (Strebel \& Orenstein, 2019). Photophobia and mild gastrointestinal symptoms have also been considered part of the disease (Drutz, 2016). Many typical clinical signs in measles infection may also be caused by other infectious diseases, including: rubella, parvovirus B19, herpes virus type 6 and dengue (Rota, et al., 2016). Measles-associated complications may be due to factors related to the host’s immune response, favouring then secondary bacterial infections; complications such as diarrhoea (8 \%), pneumonia (1 to 6\%) and otitis media (7 to 9\%) are seen. Central nervous involvement is also seen and complications such as post-infectious encephalitis (1 in every 1000 cases) and subacute sclerosing panencephalitis (1 in every 1000 cases) may occur (Rota, et al., 2016) (Strebel \& Orenstein, 2019). The complication risk augments with age, and an association has been seen in regards to higher susceptibility to complications in children with low vitamin A intake, as well as those who suffer forms of immunosuppression (Strebel \& Orenstein, 2019). Motherly-acquired IgG antibodies protect the newborn from measles; in general, these antibodies last in the newborn around 9 months. Measles is an infection related to newborns and children who live in urban environments with low vaccination coverage. Measles cases tend to predominate in scholar children (5 to 10 years of age). This average age may switch to teenagers or even adults if the vaccination coverage diminishes even more, which requires then an effort to promote immunization in older age groups (Rota, et al., 2016). 
\section{Preventive measures: vaccination.}
The high transmissibility of measles virus may be explained in terms of the high viral charges in the upper airways during the prodromal and early rash phases, in combination with the epithelial damage inducing a cough reflex. This combination turns into measles virus-aerosol generation, promoting respiratory transmission (Rota, et al., 2016). The best preventive measure against measles is vaccination; all measles vaccines contain live attenuated strains of measles virus; most of the strains derive from the Edmonston prototype strain (say, Moraten, Schwartz and Edmonston-Zagreb strains), although other wild-type-derived vaccines such as CAM-70 and Leningrad-16 exist (Moss, 2017). Measles vaccine may be administered as a combined vaccine against rubella and mumps; the application of the combined vaccine against rubella-measles provides the opportunity to diminish both rubella and congenital rubella syndrome incidence (Moss, 2017). Various studies regarding the effectiveness of the measles vaccine have encountered a high effectiveness after a dosage administered at 12 months of age or later has been performed (average effectiveness, 93\%), and even a higher effectiveness after two dosages have been administered (median effectiveness, 97\%). The WHO advises on a two-dosage schema of the measles immunization as a standard prophylaxis in measles prevention (Strebel \& Orenstein, 2019). The WHO advises on the first dosage of the vaccine to be at 9 moths of age. The ratio of children who develop a suitable measles antibody response is approximately 85\% at 9 months of age and 95\% at 12 months of age. The necessary amount of populational immunity in order to prevent measles transmission is not accomplished by means of a single dosage of the vaccine; for a second dosage should be administered. The campaigns aim for the second dosage to be administered at an age range from 5 to 15 years (Moss, 2017). The immunity induced by the vaccine most likely lasts in most of those who have been vaccinated. The MMR vaccine poses an acceptable range of secondary effects; these events include fever ($<15\%$ of the receptors), transient cutaneous rash (7 to 12 days after the vaccine administration in a 5\% of the receptors), transient lymphadenopathy (5\% in children and 20\% in adults), parotitis ($<1\%$) and aseptic meningitis (1 to 10 in a million) (Strebel \& Orenstein, 2019).  
\section{Models.}
\subsection{SIR model}
\subsubsection{Stochastic approach.}
Let us consider a compartmental model of disease, in which we have five blocks or states: susceptible, vaccinated, infected, recovered, dead:
$$
S \ \ \overrightarrow{k_1}\ \ V
$$
$$
\downarrow \ \ \swarrow
$$
$$
I \ \ \overrightarrow{k_2} \ \ R
$$
$$
\downarrow \ \ {k_4}
$$
$$
D
$$
These compartments may be interpreted as states in a discrete Markov chain, with the set of states then being, $H=\{S,V,I,R,D\}$ with a bijective function towards the set $S=\{s_i,i\in {N} \}$ such that $f:H\rightarrow S$, we have already noted that every “reaction” (with an analogy towards chemical reactions) displays a constant $k_i$, and a set of parameters, in which this set is $T=\{t_i\in {N}\}$, the stochastic matrix P for this chain is the following, 
\begin{center}
 \begin{tabular}{||c c c c c c||}
 \hline
  Variables & S & V & I & R & D \\ [0.5ex]
  \hline \hline
  S & 0.409365773 & 0.58593183 & 0.004702398 & 0 & 0 \\
  \hline
  V & 0 & 0.99999995 & $5.14851x10^{-8}$ & 0 & 0 \\
  \hline
  I & 0 & 0 & 0.2727272 & 0.63438041 & 0.0928232 \\
  \hline
  R & 0 & 0 & 0 & 1 & 0 \\
  \hline
  D & 0 & 0 & 0 & 0 & 1 \\ [1ex]
 \hline
 \end{tabular}
\end{center}
Let us set the initial vector to be, 
$$
 \pi^0 = 
 \begin {pmatrix}
 $0.4141068143$ \\
 $0.58593179$ \\
 6.82407 \times {10^{-8}}  \\
 $0$ \\ 
 $0$ 
 \end {pmatrix} ^T
$$
We may approximate the stationary distribution to be, 
$$
 \pi \approx 
 \begin {pmatrix}
 2.3359\times 10^{-249} \\
 $0.99667048$ \\
 7.05563 \times 10^{-8} \\
 $0.00290419$ \\
 $0.00042526$
 \end {pmatrix} ^T
$$
Therefore, we may realize that in the long term, the whole population will tend to be vaccinated, and only a small amount of it will become infected, however an even lesser amount of the population will still be susceptible, this due to either the effect of immunity from a previous infection or because of the vaccination status, this susceptible status in the long term appears to be even negligible. The mortality is set to be 4 in every 10,000 members of the population and the recovery rate to be 2 in every 1,000 subjects. 
Let us now compute the eigenvalues of the stochastic matrix, 
$$
\Lambda = 
 \begin {pmatrix}
 $1$ \\
 $1$ \\
 $0.99999995$ \\
 $0.409365773$ \\
 $0.2723272$
  \end {pmatrix}
  \approx 
  (P_{ii})^T
$$
That is the eigenvalues approach the transpose of the diagonal of the stochastic matrix; from this we gather that the greatest eigenvalue is $\lambda_1=\lambda_2>\lambda_3>\lambda_4>\lambda_5$; where the spectral radius $r=\lambda_1=\lambda_2=1$ is associated with two eigenvectors, say u and v,
$$
u=
 \begin {pmatrix}
  1.016911685 \times 10^{-3} \\
  $0$ \\
  $0.1277269272$ \\
  $0$ \\
  $1$ 
 \end {pmatrix}
$$
In this case, the leader vertex or state is the D state, followed by the I state, let us continue with the v eigenvector,
$$
v=
\begin {pmatrix}
6.944695232 \times 10^{-3} \\
$0$ \\
$0.8722729765$ \\
$1$ \\
$0$ 
\end {pmatrix}
$$
Now, the leader vertex is the R state followed by the I state; being the two leader vertices R and D and the second leader-vertex being in both of the cases the I state. 
\subsubsection{Non-stochastic approach.}
Alternatively, we may consider a non-stochastic approach; let us consider the following system of equations,
$$
-\frac{dS}{dt} = - \frac{d(S_0 - x -y)}{dt}
$$
$$
=\frac{dx}{dt}+\frac{dy}{dt}
$$
$$
=k_1\left(S_0-x\right)+k_2\left(S_0-y\right)
$$
Which describes the formation of both vaccinated and infected populations from the susceptible set. Now, 
$$
-\frac{dV}{dt}=-\frac{d\left(x-y\right)}{dt}
$$
$$
=-\frac{dx}{dt}+\frac{dy}{dt}
$$
$$
=k_3\left(x-y\right)
$$
Describes the rate of formation of infected people from the vaccinated set. Finally,
$$
-\frac{dI}{dt}=-\frac{d\left(x+y-u-z\right)}{dt}
$$
$$
=\frac{du}{dt}+\frac{dz}{dt}-\frac{dx}{dt}-\frac{dy}{dt}
$$
$$
=k_3\left(x+y-u\right)+k_4\left(x+y-z\right)
$$
And,
$$
-\frac{dR}{dt}=-\frac{du}{dt}=1
$$
$$
-\frac{dD}{dt}=-\frac{dz}{dt}=1
$$
We may group them,
$$
x^\prime+y^\prime=S_0\left(k_1+k_2\right)-k_1x-k_2y
$$
$$
-x^\prime+y^\prime=k_3x-k_3y
$$
$$
u^\prime+z^\prime-x^\prime-y^\prime=k_3\left(x+y-u\right)+k_4\left(x+y-z\right)
$$
$$
-u^\prime=1
$$
$$
-z^\prime=1
$$
For which we may define a matrix Y,
$$
Y=
\begin{pmatrix}
 x \\
 y \\
 u \\
 z \\
 1
\end{pmatrix}
$$
of the variables involved in the process and a constant; a matrix A,
$$
A=
\begin{pmatrix}
 -k_1 & -k_2 & 0 & 0 & S_0 (k_1 + k_2 ) \\
 k_3 & -k_3 & 0 & 0 & 0 \\
 k_3 + k_4  & k_3 + k_4  & -k_3 & -k_4 & 0 \\
 0 & 0 & 0 & 0 & 1 \\
 0 & 0 & 0 & 0 & 1
\end{pmatrix}
$$
And a matrix $Y^\prime$,
$$
Y^\prime = 
\begin {pmatrix}
x^\prime \\
y^\prime \\
u^\prime \\
z^\prime 
\end {pmatrix}
$$
Then would we have a system of differential equations,
$$
AY=Y^\prime
$$
This linear system has a solution (Zill, 2009),
$$
Y\left(t\right)=\sum_{i=1}^{n}{e^{\lambda_it}V_i}
$$
For which we shall define the constants ($k_1$ stands for the rate of the ‘real’ vaccinated individuals in the susceptible population, this done with the data from the National Mexican Committee of Vaccination which reported vaccination rates from 1990 to 2000, due to the fact that the second dosage tends to happen at the second year of age, the data presented is only the one from the dosages administered, independent of the fact whether it is the first or the second dosage, due to this, we computed the real vaccination rate from 1995 forth when we know the actual amount of vaccinated people, who have already received the second dosage; for this constant we have also considered the fact that geographic areas with less than or equal to 1500 inhabitants were not taken into account in the vaccination schema, then we subtracted the amount of those born in rural areas in those periods; therefore we may consider this constant to be the rate of those who have received the second dosage of the vaccine who lived in geographic areas of more than 1500 inhabitants. $k_2$ stands for the rate of those infected from the susceptible pool. $k_3$ stands for the rate of recovered patients from the infected pool. $k_4$ stands for the rate of mortality from infected patients. $k_5$ stands for the rate of infected from the vaccinated pool. We have also considered the susceptible population to be $146.54x10^6$ following the exponential rule of population applied to Mexico, this is because there is no data available (yet) from the 2020 census),
\begin{center}
 \begin{tabular}{||c c||}
 \hline
  Constant & $Value \pm SD $ \\ [0.5ex]
  \hline \hline
  $k_1$ & $0.585931829 \pm 0.07229194$ \\
  \hline
  $k_2$ & $0.004702398 \pm 0.000218881$ \\
  \hline
  $k_3$ & $0.634380406 \pm 0.4407335$ \\
  \hline
  $k_4$ & $0.092892322 \pm 0.176686532$ \\
  \hline
  $k_5$ & $5.14851x10^{-8} \pm 1.07708x10^{-7}$ \\ [1ex]
 \hline
 \end{tabular}
\end{center}
In the former equation, $\lambda_i$ is the i-th eigenvalue of the A matrix and $V_i$ the i-th eigenvector associated with said eigenvalue; in this case, we obtain the eigenvalues of the matrix for the values of the matrix $k_i=k_i$ (no standard deviation involved),
$$
\Lambda_1=
\begin{pmatrix}
0 \\
1 \\
-0.634380406 \\ 
-0.6699049694 \\ 
-0.5504072656
\end{pmatrix}
$$
The eigenvectors for given eigenvalues are shown,
\begin{center}
 \begin{tabular}{||c c c c c c||}
 \hline
  Var-EV & $v_1$ & $v_2$ & $v_3$ & $v_4$ & $v_5$ \\ [0.5ex]
  \hline \hline
  x & 0 & 54637446.31 & 0 & -0.0289759648 & 0.01349724863 \\
  \hline
  y & 0 & 21207379.41 & 0 & -0.1019658193 & 0.1019658193 \\
  \hline
  u & -0.146429987 & 33749715.19 & 1 & 1 & 1 \\
  \hline
  z & 1 & 1 & 0 & 0 & 0 \\
  \hline
  1 & 0 & 1 & 0 & 0 & 0 \\ [1ex]
 \hline
 \end{tabular}
\end{center}
Thus, displaying the solution,
$$
Y_1\left(t\right)=v_1+e^tv_2+e^{-0.634380406t}v_3+e^{-0.6699049694t}v_4+e^{-0.5504072656t}v_5
$$
We show now the eigenvalues for the cases $k_i\pm sd$,
\begin{center}
 \begin{tabular}{||c c||}
 \hline
  $k_i \pm sd$ & $k_i - sd$ \\ [0.5ex]
  \hline \hline
  -1.086791796 & 1 \\
  \hline
  -1.07445376 & -0.510888052 \\
  \hline
  1& -0.197058898 \\
  \hline
  -0.6458857338 & -0.19430706 \\
  \hline
  0 & 0  \\ [1ex]
 \hline
 \end{tabular}
\end{center}
The eigenvector for $k_i - sd$,
\begin{center}
 \begin{tabular}{||c c c c c c||}
 \hline
  Var-EV & $v_1$ & $v_2$ & $v_3$ & $v_4$ & $v_5$ \\ [0.5ex]
  \hline \hline
  x & 50136913.64 & -7.416910083 & $3.577164711x10^{-4}$ & $0$ & $0$ \\
  \hline
  y & $8156994.64$ & $4.552256859$ & $-0.02525833169$ & $0$ & $0$ \\
  \hline
  u & 5394111.985 & 1 & 1 & 1 & 0.4312463479 \\
  \hline
  z & 1 & 0 & 0 & 0 & 1 \\
  \hline
  1 & 1 & 0 & 0 & 0 & 0  \\ [1ex]
 \hline
 \end{tabular}
\end{center}
And the one for $k_i + sd$,
\begin{center}
 \begin{tabular}{||c c c c c c||}
 \hline
  Var-EV & $v_1$ & $v_2$ & $v_3$ & $v_4$ & $v_5$ \\ [0.5ex]
  \hline \hline
  x & $1.066376173x10^{-4}$ & 0 & 58693455.2 & 0.09092105197 & 0 \\
  \hline
  y & $-9.286501305x10^{-3}$ & $0$ & $30400004.5$ & $0.2279462307$ & $0$ \\
  \hline
  u & 1 & 1 & 57723395.53 & 1 & -0.2508985124 \\
  \hline
  z & 0 & 0 & 1 & 0 & 1 \\
  \hline
  1 & 0 & 0 & 1 & 0 & 0  \\ [1ex]
 \hline
 \end{tabular}
\end{center}
Such that we obtain the solutions,
$$
Y(t)=
\begin{pmatrix}
k_i\rightarrow v_1+e^tv_2+e^{-0.634380406t}v_3+e^{-0.6699049694t}v_4+e^{-0.5504072656t}v_5 \\
k_i-sd\rightarrow e^tv_1+e^{-0.510888052t}v_2+e^{-0.197058898t}v_3+e^{-0.19430706t}v_4+v_5 \\
k_i+sd\rightarrow e^{-1.086791796t}v_1+e^{-1.07445376t}v_2+e^tv_3+e^{-0.6458857338t}v_4+v_5
\end{pmatrix}
$$
\subsection{Cellular compartmentalization. }
SLAMF1, Nectin-4 and CD46 show specific cellular expression patterns, thereof we may take advantage of this in order to compute the temporal distribution of viral tropism in the host. Let us call the existence of expression of either protein a condition $c_i$ and the set of possible cells with the condition $c_i$, $V=\left\{l_1,l_2,\ldots,l_n\right\}$; we postulate the following,
$$
\forall l_n\in V\exists R|c_i\in\left(l_j\land l_k\right)\rightarrow\left(l_jRl_k\land l_kRl_j\right)
$$
Such that two cells possessing the same condition will exhibit a relationship, this feature builds up to a graph. This graph G is defined by $G=\left(V,E\right)$ where V is the set of vertices and E is a set of edges; the graph has an adjacency matrix M (Fig. 1) with the adjacency condition, 
$$
m_{ij}=\left(C=\left\{c_i\right\}\right),\left|C\right|\in l_n\rightarrow\left|C\right|
$$
\begin{figure}[htp]
    \centering
    \includegraphics[width=12cm]{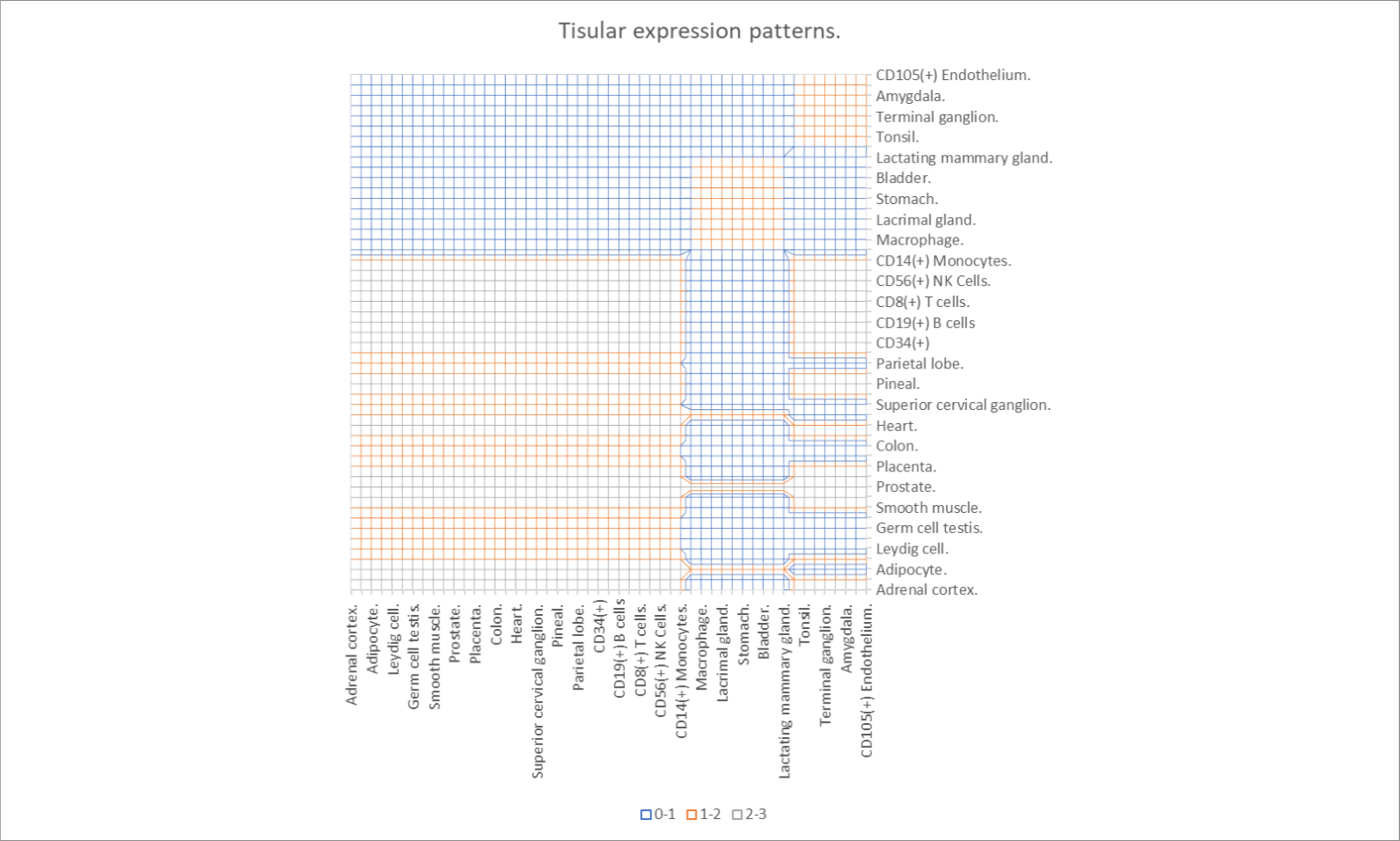}
    \caption{Graph exhibiting the adjacency matrix of the tisular/cellular expression of SLAMF1, nectin 4 and CD46.}
    \label{fig:matrix_cell}
\end{figure}

The ij-th member of the matrix of adjacency shall be equal to the cardinality of the set of conditions the specific cell expresses. We are going to approximate the eigenvalues of this matrix by means of the Gershgorin circle theorem (Fig. 2) (Weisstein, 2020), thereof we possess a matrix M with $m_{ij}\in {C}$, and for each i, we define the disc, 
$$
D_i=\left\{z\in {C}:\left|z-m_{ii}\right|\le\sum_{j\neq i}\left|m_{ij}\right|\right\}
$$

\begin{figure}[htp]
    \centering
    \includegraphics[width=12cm]{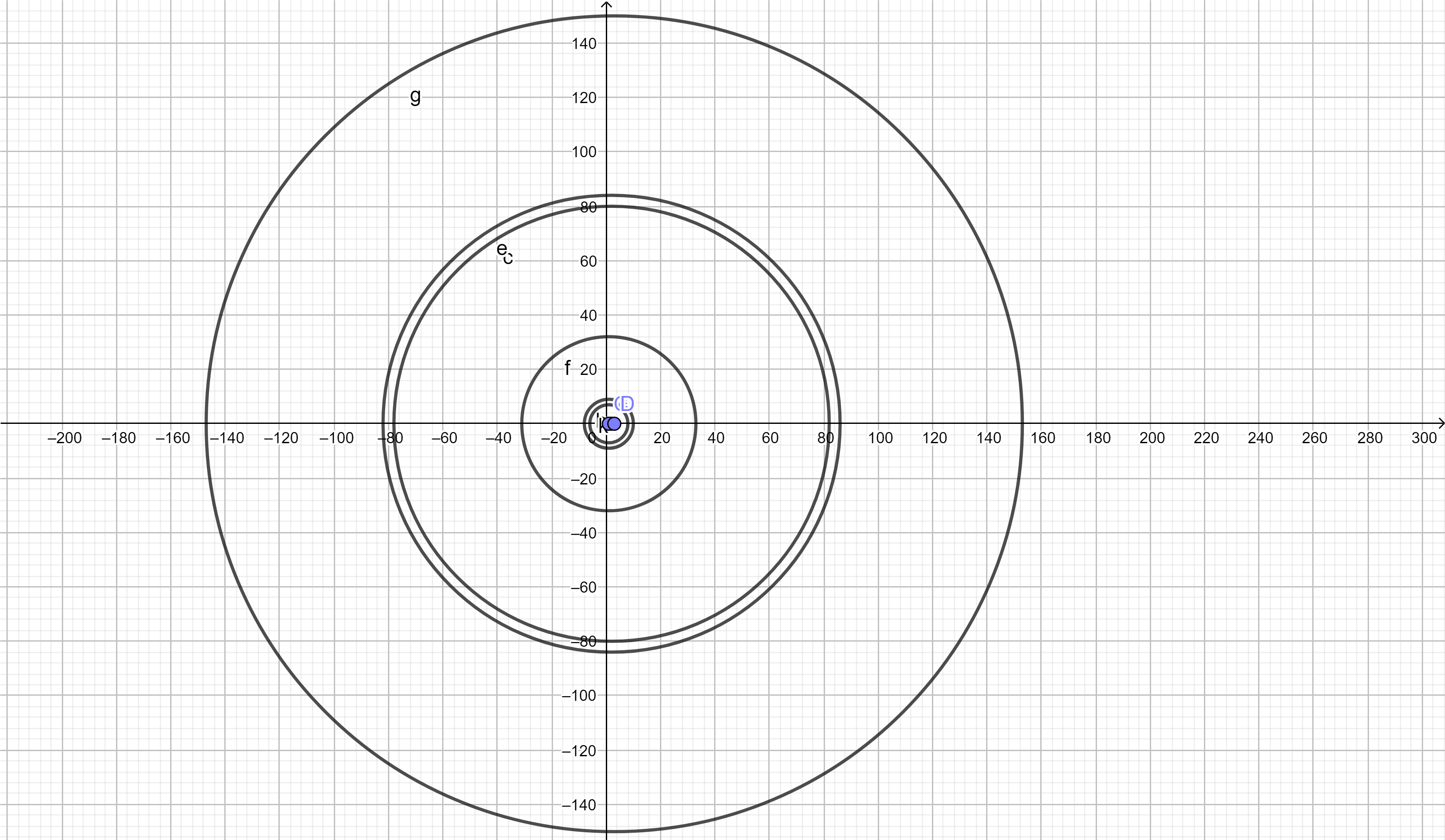}
    \caption{Graphic representation of the Gershgorin theorem applied to the approximation of the eigenvalues corresponding to the adjacency matrix of the cellular compartmentalization of measles virus interactions in the complex plane.}
    \label{fig:gersh}
\end{figure}

That is, the radius of the disc is lesser than the sum of all the non-diagonal entries of the matrix, and the eigenvalues (the matrix of eigenvalues is defined as $\Lambda$) of M lie in,
$$
\Lambda\subset\bigcup_{i=1}^{n}D_i
$$
Should we consider $\Lambda\subset {R}$ then would we obtain $\Lambda\in\left[-147,153\right]$ that is, that our real eigenvalues lie in that range (by observation of the adjacency matrix we realize that $\max{\sum_{i,j}\left|m_{ij}\right|}=150$ corresponding to the node of prostate, therefore, this may lead our thought to the possible fact that prostate is the leader vertex of the network and that the radial spectra r of the network is $r=max \lambda \leq 153$). 
When we convert our adjacency matrix to a stochastic one, we learn then from the stationary distribution of the corresponding Markov chain that the virus shall tend to remain in: lymph node, tonsil, appendix, terminal ganglion, retina, amygdala, spinal cord and CD105(+) endothelium (Fig. 3) which may clinically correlate to advanced stages of disease such as encephalitis (terminal ganglion) which has been reported to be one of the most common complications of measles virus infection (Fisher, Defres, \& Solomon, 2014), other not so common complications include retinitis (Neppert, 1994), tonsillitis (Lai, Lin, Wang, \& Chen, 2017) and induction of giant cells observed in appendix promoting then appendicitis (Whalen, Klos, Kovalcik, \& Cross, 1980). 
\begin{figure}[htp]
    \centering
    \includegraphics[width=12cm]{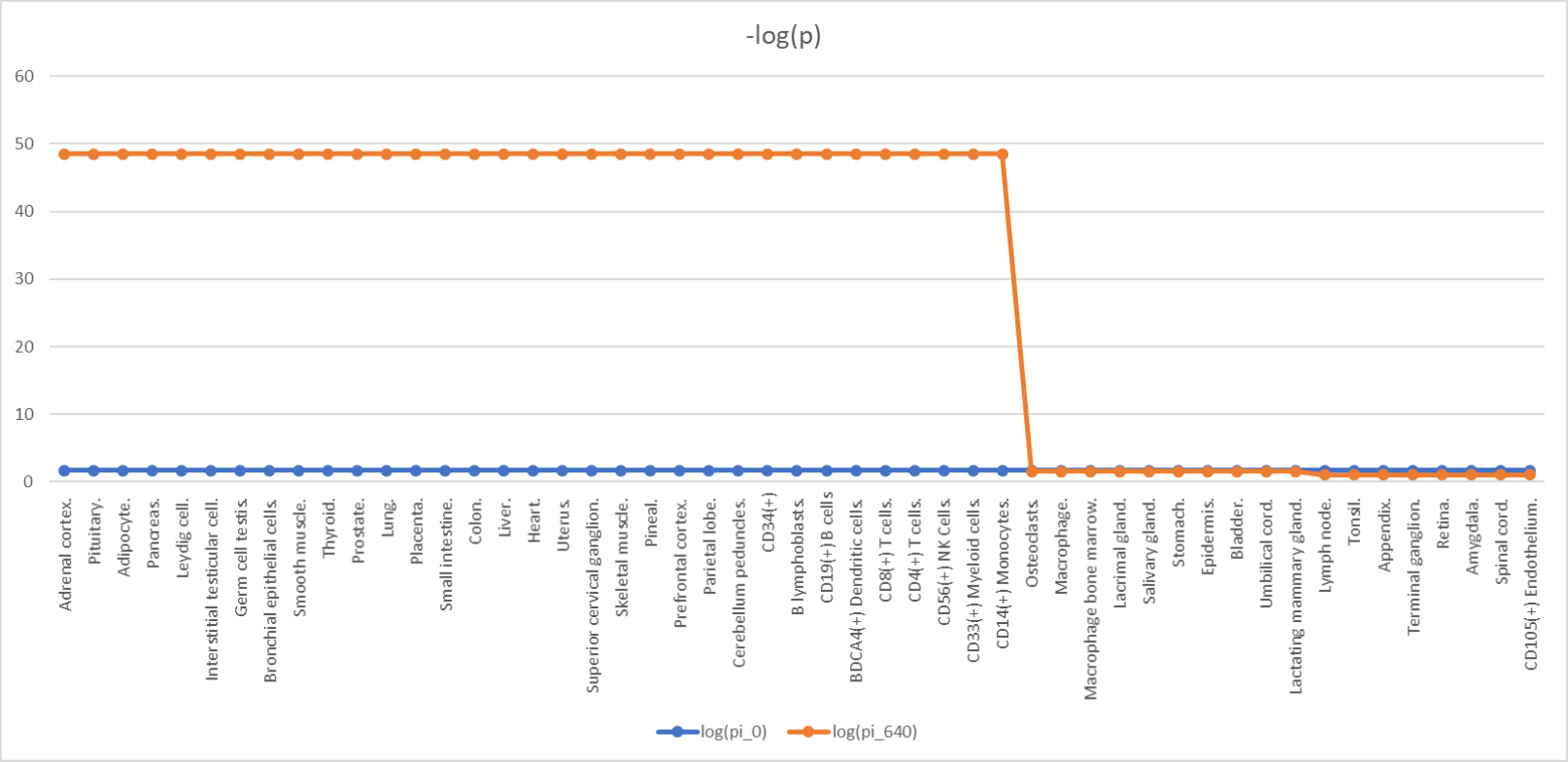}
    \caption{Graph showing the $-\log_{10}{\left(p_i\right)}$ where $p_i$ refers to the i-th probability. }
    \label{fig:cellul}
\end{figure}
\subsection{Geographic distribution tendencies}
\subsubsection{Analysis of the Mexico City outbreak.}
Let us consider Mexico City and its boroughs as elements of a network in which certain rates of individuals leave one borough and enter another, whereas those from the other borough leave this one and enter the first one, there exists a ‘reversible’ process however with obvious different rates; these allows us to build a discrete model of the large-scale population behaviour in Mexico City. For these rates we will consider data obtained from the subway system of the city as well from rates of traffic from main avenues. Due to the fact that the subway system of Mexico City only reports the affluency of each subway station, we need to know how many people enter to every borough from another borough or remain in the same borough, therefore let us consider the following schema,
$$
A\rightleftharpoons B
$$
Where the rate of A towards B is $k_1$ and the rate of B towards A is $k_2$; in order to determine how many people stay in a borough or leave it, we will consider that in regards to the subway movements, the amount of people who come down from a specific set of stations within a borough is proportional to the ratio of the population of that borough, 
$$
k_i\ \alpha\ r
$$
where r stands for the ratio $r=\frac {P_b}{P_t}\ $ where $P_b$ stands for the population of the borough and $P_t$ for the total amount of population, in this model we consider the populations of the states of Mexico and Morelos as if they were ‘boroughs’, in order to take in account the very important interactions Mexico City poses with these states (Fig. 4). 

\begin{figure}[htp]
    \centering
    \includegraphics[width=12cm]{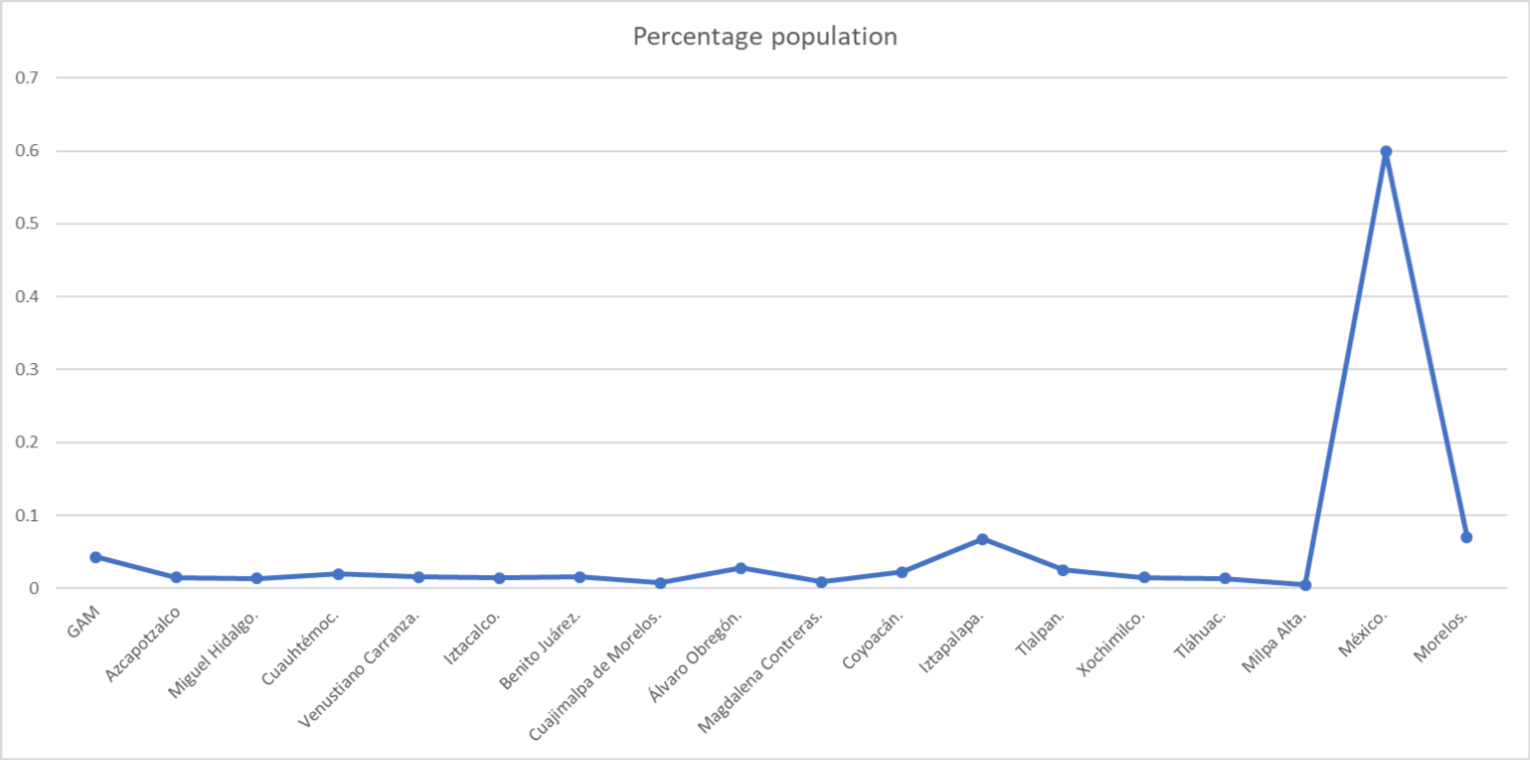}
    \caption{Population ratios of each borough comprising Mexico City and metropolitan area. }
    \label{fig:perc}
\end{figure}

We may imagine this model as a balloon being pinched in the both ends (representing a borough being pierced by a subway line) with some people already in the balloon, some entering by means of the initial pierce, and some more leaving towards the second pierce, however we ought to account for all of the ‘pinches’ this balloon endures, therefore, we need to know all the affluency towards this borough, the affluency original of this borough and the effluency from this borough; therefore some people will stay in the borough and some other will leave, the way in which we shall know how many people will stay in every borough is by means of the ratio r we have already commented. Let us provide an example, the Gustavo Adolfo Madero borough is ‘pierced’ by five subway lines, some originating in these borough, some others do not, Line B comes from Mexico State, therefore we compute the affluency $A_m$ of those stations from Line B located at Mexico State, then, some people from this affluency will stay in Mexico State while some others will pierce towards Gustavo Adolfo Madero (GAM) borough, the amount of those who stay at Mexico State are $s_m=r_m \bullet A_m$ that is the stay rate equals the ratio of population Mexico State possesses $r_m$ times the affluency from the stations; while those who enter are then $e=A_m-s_m$, in here this entry population mixes with the affluency of those who enter the subway stations at the GAM borough with an affluency $A_G$, now some people will leave the borough and some others will stay, the amount of those who shall stay is $s_G=r_G\times A_G$ thus the amount of those who leave is $e=A_G-s_G$; this becomes quite easily an algorithm, in which we compute the affluency from the root borough, those who stay, those who leave, those who enter into a new borough, and so forth. At the end of the algorithm, we obtain valuable quantities, those are the total amount of people who are found at the borough at a given time, let us call this quantity $l$ the entry plus affluency, $l=e_i+A_i$, this allows us not to count twice a person and to normalize our stochastic process. Then, we sum the quantities l for each borough, followed by a description of those lines that pierce from the balloon outwards, these being outflows or rates of outside flow, if the borough is, for instance, GAM, then line B pierces out towards Cuauhtémoc, then the ratio of the exit flow and the sum of the l quantities of the borough GAM will provide us the actual rate of flow $\phi$ from one borough to the other,
$$
\phi=\frac{e_i}{\sum(l_i)}
$$
This rates of flow for each borough have a feature, 
$$
\sum_{i=1}^{n} {\phi_i}=1
$$
Then we realize these rates of flow are indeed transition probabilities, and members of rows from a random matrix, in this case, rows from a stochastic matrix (Fig. 5) corresponding to the Markov process of Mexico City inhabitants’ movements by means of the subway affluencies. The states of this Markov process are the boroughs $S=\{s_i\}$ and the temporal parameter is a discrete one. We are in position now to describe the Mexico City outbreak, which we will represent by means of the initial vector of the Markov process $\pi^0$ whose probabilities are given by a frequentist view of the cases in Mexico City, that is, $p_i=\frac{c}{n}$ the probability of transition $p_i$ equals the ratio of the amount of cases in the i-th borough and the total number of cases n (Fig. 6). When we compute the distribution of probabilities in different moments, we obtain a graph which is shown in Fig. 6 in which we may regard that in time, the areas less affected by measles will tend to be Cuajimalpa ($\pi_C=1.024x10^{-194}$), Magdalena Contreras ($\pi_{MC}=5.9876x10^{-40}$), Tlalpan, Xochimilco, Milpa Alta and Morelos with similar rates; the areas with the higher likelihood to present cases or to concentrate more cases are Álvaro Obregón with $\pi_{AO}=0.207358185$ and State of Mexico with $\pi_{SM}=0.20140335$.

\begin{figure}[htp]
    \centering
    \includegraphics[width=12cm]{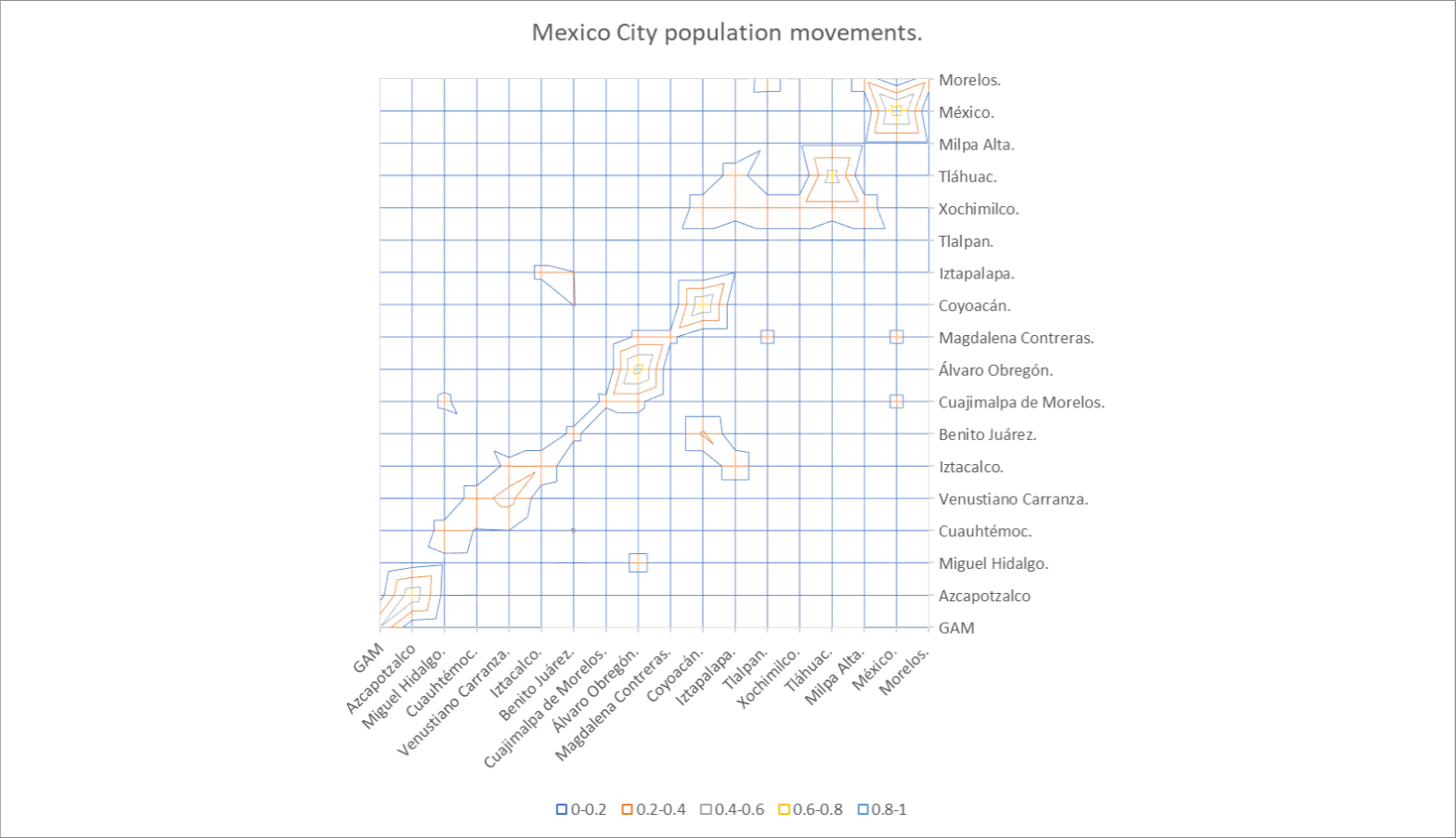}
    \caption{Stochastic matrix of human movements in Mexico City. }
    \label{fig:cdmx}
\end{figure}

\begin{figure}[htp]
    \centering
    \includegraphics[width=12cm]{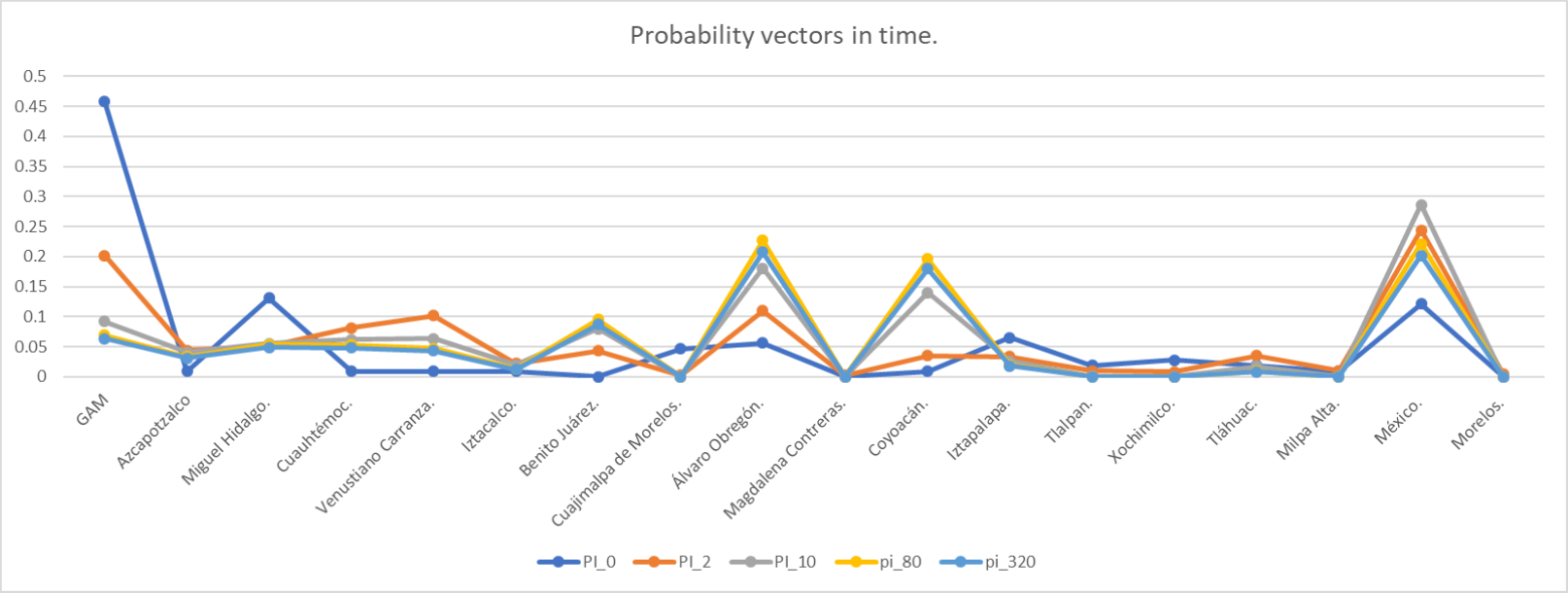}
    \caption{Evolution of probabilities distribution of measles cases in different moments of the temporal parameter.  }
    \label{fig:vect}
\end{figure}

\subsubsection{Analysis of global outbreak.}
We present a Markov chain in which the discrete states are the countries worldwide $S=\{s_i\}$ and the temporal parameter is one for which the dimension remains to be determined, however the time is discrete; the existence of a probability of transition $p_ij$ between one i-th country and a j-th is by means of a physical political boundary sharing or a connection through a commercial flight, the stochastic matrix is shown in Fig. 7. 

\begin{figure}[htp]
    \centering
    \includegraphics[width=12cm]{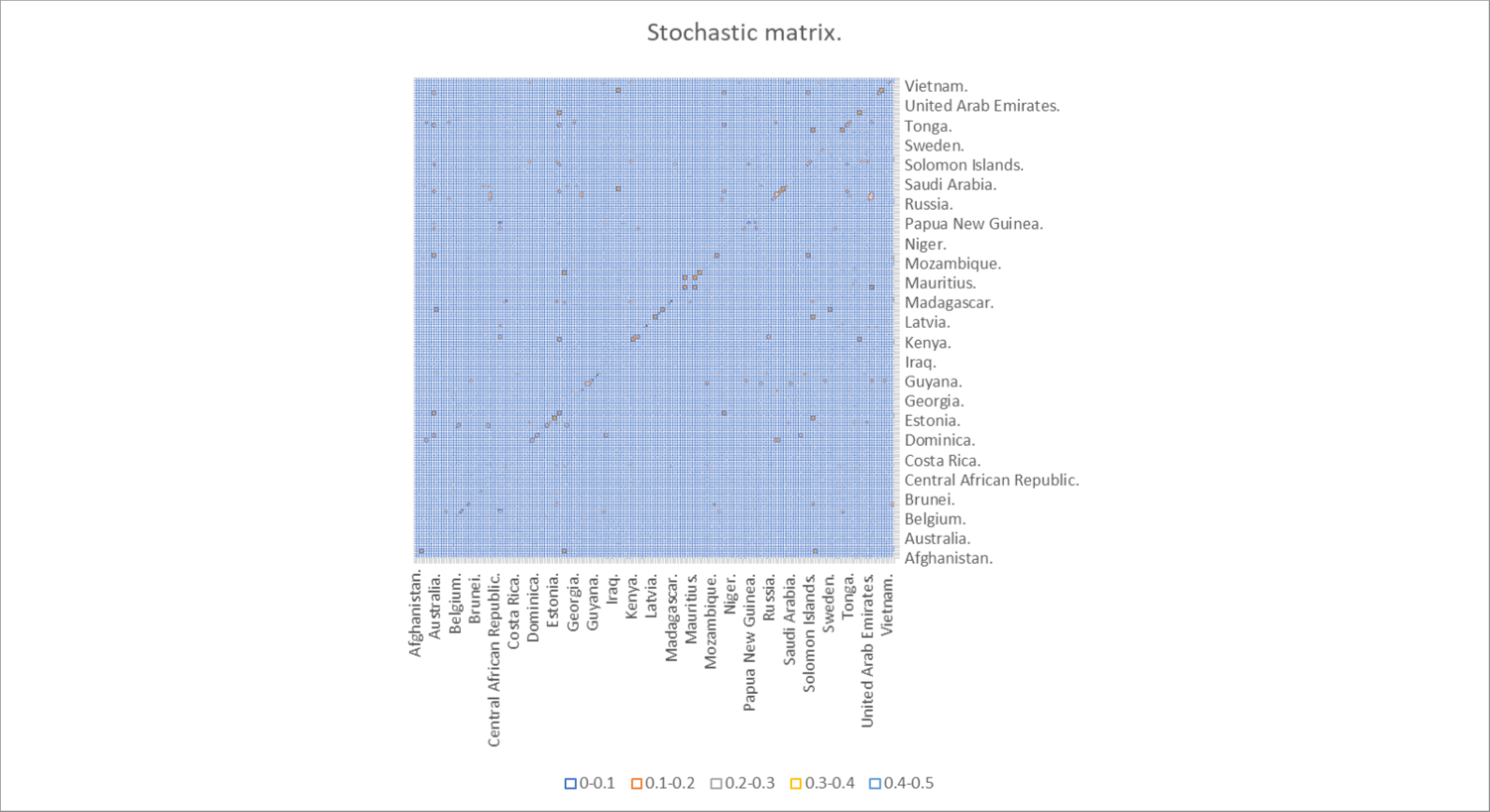}
    \caption{Stochastic matrix displaying the interactions nations have amongst them in terms of physical interactions (political territory boundaries) and air trafficking.  }
    \label{fig:nation}
\end{figure}

The initial vector corresponds to the amount of global cases worldwide, where $\pi_i=\frac{c_i}{c_t}\ $ and $c_i$ stands for the amount of cases of measles in that i country and $c_t$ stands for the total amount of cases; the quantities of cases were obtained from the WHO reports from 2019 and the Mexican Health Ministry from 2020, therefore, our initial vector stands for that temporal unit between January 2019 and March 2020. By computing the stationary distribution, are we able now to predict a long-term distribution of the cases worldwide, taking in account the various air traffic influences in the transmissability of measles; this stochastic matrix could be of use in many other infectious diseases or even modern patterns of migration. When we compute the probability vectors in different moments in time, we obtain the Fig. 8 in which there exists a high homogenization in the distribution of cases worldwide, however when we compare the vectors when $n=0\wedge n=128$ (Fig. 9) we observe that those countries that started with p=0 of cases, they preserve such probability; the countries with the highest concentration of cases are those with the highest rate of interactions with other countries: France ($p=0.02328$), Turkey ($p=0.0228462$), South Africa ($p=0.0198728$), United Kingdom ($p=0.0198559$); being then the European continent the one with the highest contribution to the cases; Mexico shows a contribution of 0.723\% of the cases worldwide. Those countries with $p=0$ are: East Timor, Eswatini, Kirbiati, Marshall Islands, Federated States of Micronesia, Nauru, Seychelles and Tuvalu. 

\begin{figure}[htp]
    \centering
    \includegraphics[width=12cm]{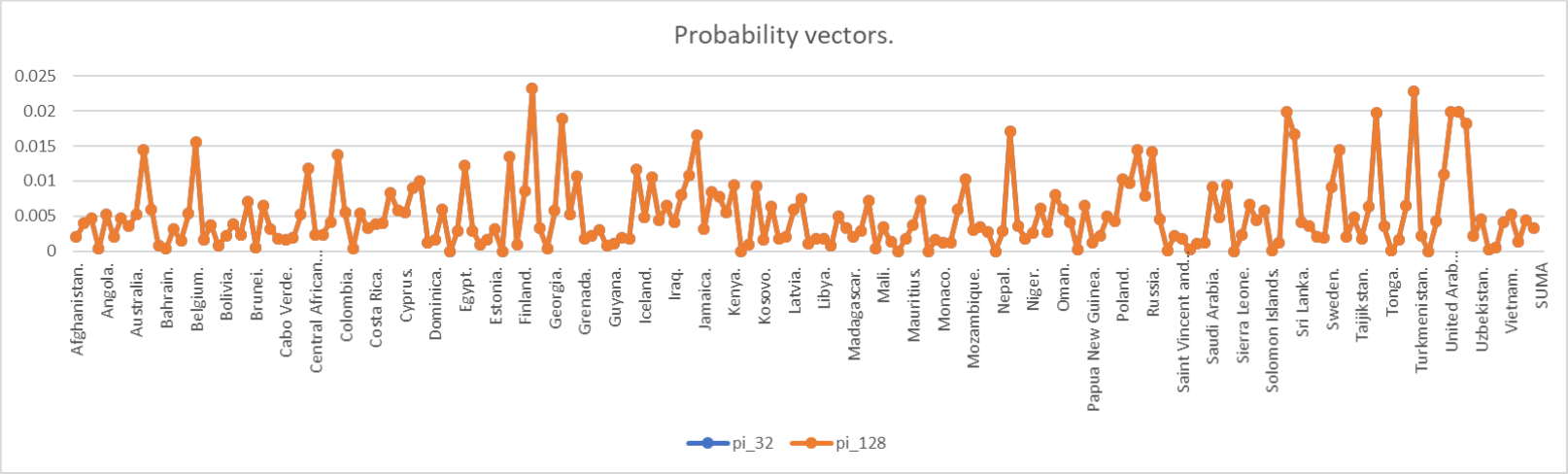}
    \caption{Probability vectors when $n=32\wedge n=128$, we can tell of a much smoother homogenization in the distribution of probabilities.  }
    \label{fig:probvec}
\end{figure}

\begin{figure}[htp]
    \centering
    \includegraphics[width=12cm]{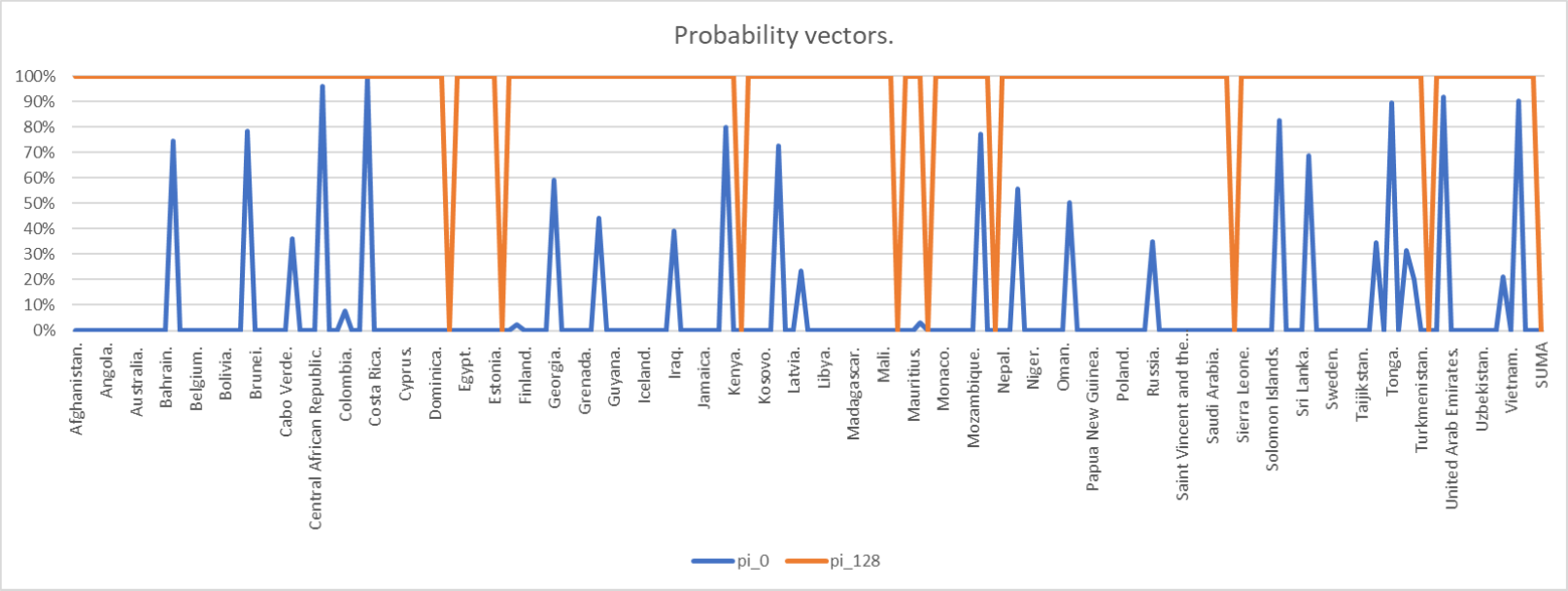}
    \caption{Distribution of probabilities vectors corresponding to cases worldwide, comparing $n=0\wedge n=128$.  }
    \label{fig:pvec2}
\end{figure}

\subsubsection{Genotype tendencies at the global scale.}
As previously discussed, the measles virus shows diversity by means of changes to its H protein, in which we may note several genotypes of the virus, these genotypes are time- and space-dependent, and we will describe these genotype fluctuations in a time-dependent manner in their geographic distribution using the adjacency matrix devised in the subsection ‘Analysis of global outbreak’ in order to promote preparedness in countries in regards to the idiosyncrasies of infection of each genotype, and development of more directed vaccines or therapies. The adjacency matrix of the countries will be converted unto a stochastic one, the Markov chain associated to the stochastic matric shows a set of states $S=\{s_i\}$ in which every $s_i$ state is a country, T is the temporal parameter and in this case the units are not defined, and the temporal parameter set is $T=\{t_i\ \in {N}\}$ such that the time parameter is a discrete one. Every genotype will display an initial vector and will be an independent Markov process; these initial vectors are shown in Fig. 10. When the number of iterations n=8, then do we obtain Fig. 11, where we regard that for those initial $pi_i^0\neq 0$, there exist the following distributions, e.g.: Angola shows $max(\pi_{An}^8)=0.007393$ corresponding to B2, Austria shows $max(\pi_{Au}^8)=0.0151245$ corresponding to D6, Republic of the Congo shows a maximum of $p=0.0047079$ corresponding to B2, Denmark shows a maximum of $p=0101149$ corresponding  to D4; for example, in the case of Mexico, we know from the information from the Ministry of Health (SALUD, 2020) that the D8 genotype is the one prevalent in the 2020 measles outbreak, and the $max(\pi_{Mex}^8)=0.00073567$ which corresponds indeed to the D8 genotype, from this we may gather that air traffic and political limits between countries are indeed determinants in measles virus’ genotypes spread and distribution worldwide; therefore there exist transitions between one initial state towards a different one in time; the observation of these changes is a priori, should we want to know whether migration, air traffic is a determinant of measles virus’ genotypes, we need to promote a posteriori study in order to correlate each probability. 

\begin{figure}[htp]
    \centering
    \includegraphics[width=12cm]{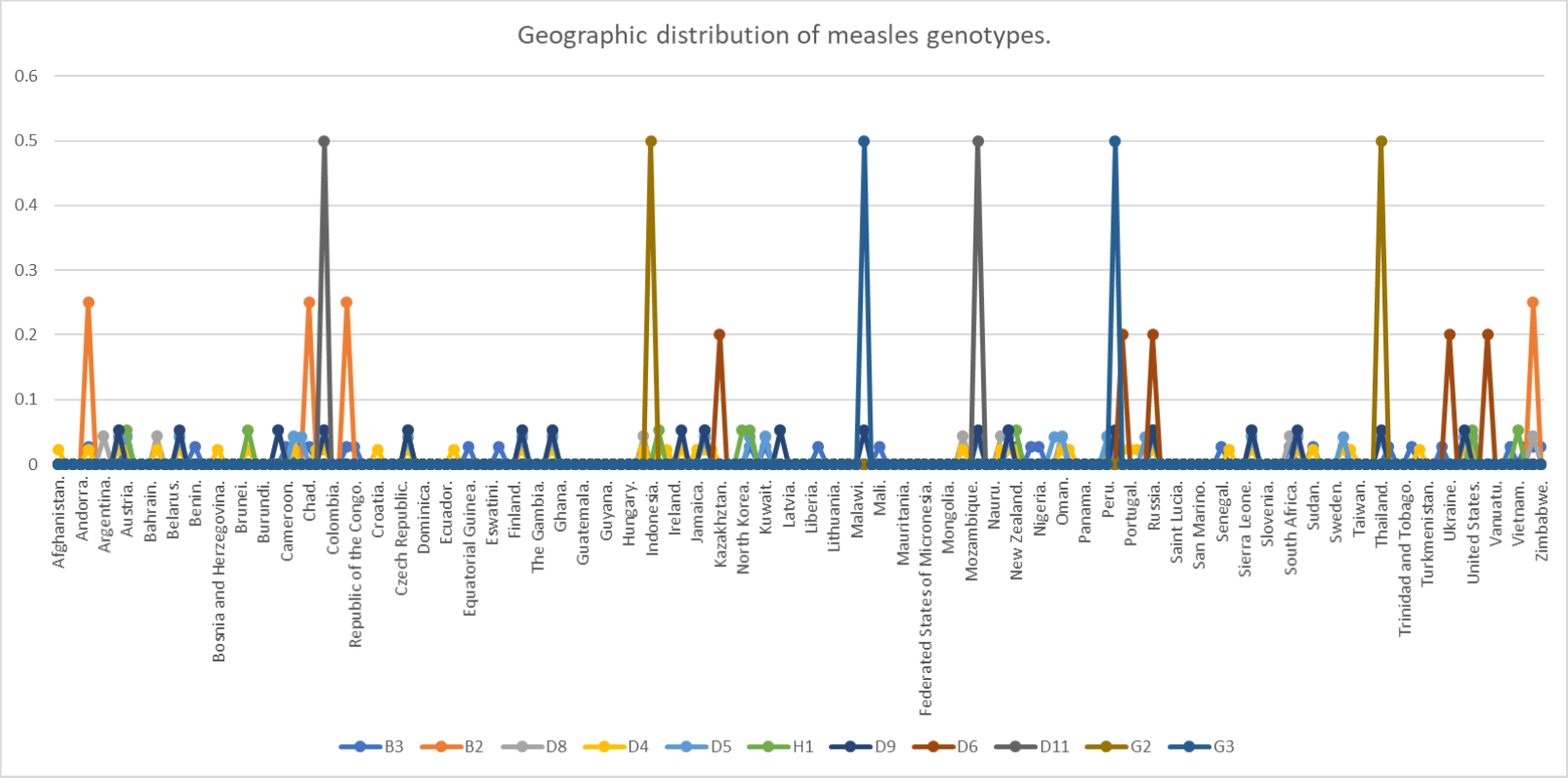}
    \caption{ Worldwide geographic distribution of measles genotypes, initial vectors for each genotype.  }
    \label{fig:genoin}
\end{figure}

\begin{figure}[htp]
    \centering
    \includegraphics[width=12cm]{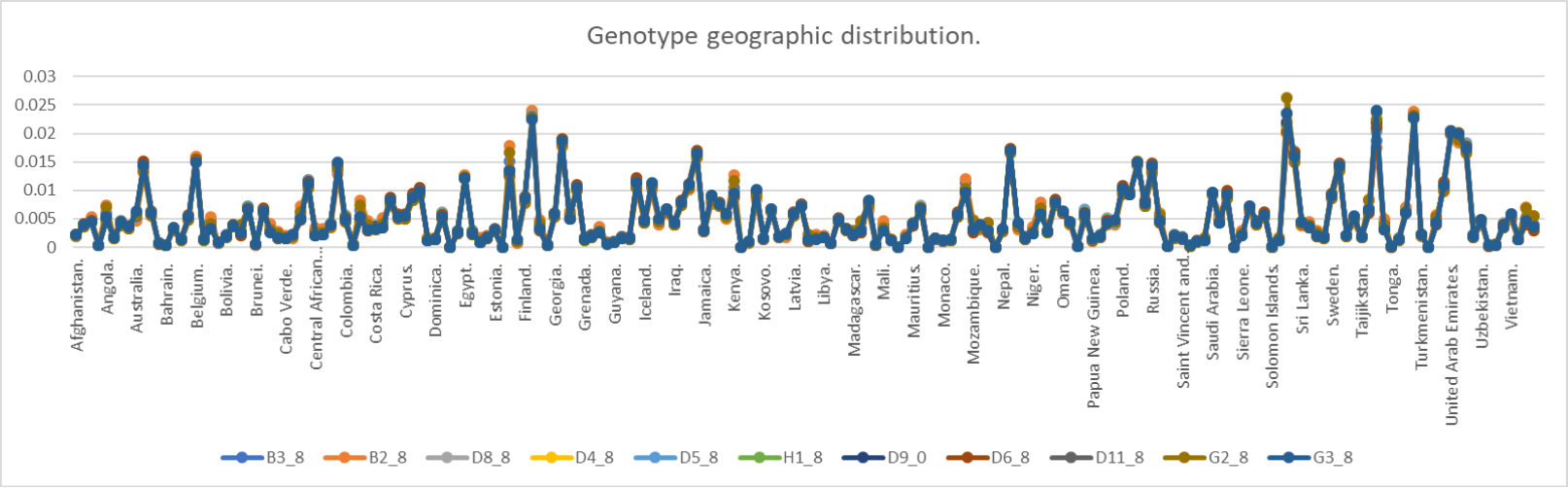}
    \caption{ Geographic distribution of measles virus genotypes when n=8.  }
    \label{fig:genot}
\end{figure}

\section{References.}
Bhattacharjee, S., Kumar Jaiswal, R., \& Kumar Yadava, P. (2019, January 31). Measles virus phosphoprotein inhibits apoptosis and enhances clonogenic and migratory properites in HeLa cells. J. Biosci., 44(10), 1-9. doi:10.1007/s12038-018-9834-6

Bloyet, L.-M., Brunel, J., Dosnon, M., Hamon, V., Erales, J., Gruet, A., Gerlier, D. (2016, December 9). Modulation of re-initiation of measles virus transcription at intergenic regions by Pxd to N tail binding strength. PLoS Pathogens., 12(12), 1-39. doi:10.1371/journal.ppat.1006058

Bologna, C., Buonincontri, R., Serra, S., Vaisitti, T., Audrito, V., Brusa, D., Deaglio, S. (2016, January). SLAMF1 regulation of chemotaxis and autophagy determines CLL patient response. The Journal of Clinical Investigation., 126(1), 181-195. doi:10.1172/JCI83013

Broz, P., \& Dixit, V. (2016). Inflammasomes: mechanism of assembly, regulation and signalling. Nature Reviews | Immunology., 16, 407-421. doi:10.1038/nri.2016.58

Cassandri, M., Smirnov, A., Novelli, F., Pitolli, C., Agostini, M., Malewicz, M., Raschellà, G. (2017). Zinc-finger proteins in health and disease. Cell Death Discovery, 3(17071). doi:10.1038/cddiscovery.2017.71

Challita-Eid, P., Satpayev, D., Yang, P., An, Z., Morrison, K., Shostak, Y., Stover, D. (2016). Enfotumab vedotin antibody-drug conjugate targeting nectin-4 is a highly potent therapeutic agent in multiple preclinical cancer models. Cancer Research., 1-33. doi:10.1158/0008-5472.CAN-15-1313

Chung, D. (2001, July). Measles vaccine virus fights lymphoma. The Lancet Oncology, 2, 397. doi:10.1016/S1470-2045(00)00410-1

Detre, C., Keszei, M., Romero, X., Tsokos, G., \& Terhorst, C. (2010, June). SLAM family receptors and the SLAM-associated protein (SAP) modulate T cell functions. . Semin. Immunopathol., 32(2), 157-171. doi:10.1007/s00281-009-0193-0

Dias Junior, A., Sampaio, N., \& Rehwinkel, J. (2018). A balancing act: MDA5 in antiviral immunity and autoinflammation. Cell Press., 1608, 1-11. 

doi:10.1016/j.tim.2018.08.007

Dong, Y.-D., Yuan, Y.-L., Yu, H.-B., Tian, G.-J., \& Li, D.-Y. (2019). SHCBP1 is a novel target and exhibits tumor-promoting effects in gastric cancer. Oncology Reports., 41, 1649-1657. doi:10.3892/or.2018.6952

Drutz, J. (2016). Measles. Pediatrics in Review., 37(5), 220-221. doi:10.1542/pir.2015-0117

Filipovich, A., Zhang, K., Snow, A., \& Marsh, R. (2010, July 26). X-linked lymphoproliferative syndromes: brothers or distant cousins? Blood, 116(18), 3398-3408. doi:10.1182/blood-2010-03-275909

Fisher, D., Defres, S., \& Solomon, T. (2014, May 26). Measles-induced encephalitis. . Q. J. Med., 1-6. doi:10.1093/qjmed/hcu113

Gopalkrishna Pai, S., Carneiro, B., Moya, J., Costa, R., Abner Leite, C., Barroso-Sousa, R., Giles, F. (2017). Wnt/beta-catenin pathway: modulating anticancer immune response. Journal of Hematology \& Oncology., 10(101), 1-12. doi:10.1186/s13045-017-0471-6

Gordiienko, I., Shlapatska, L., Kovalevska, L., \& Sidorenko, S. (2019). SLAMF1/CD150 in hematologic malignancies: silent marker or active player? Clinical Immunology, 204, 14-22. doi:10.1016/j.clim.2018.10.015

Guryanov, S., Liljeeros, L., Kasaragod, P., Kajander, T., \& Butcher, S. (2016, March). Crystal structure of the measles virus nucleoprotein core in compex with an N-terminal region of phosphoprotein. Journal of Virology. , 90(6), 2849-2858. doi:10.1128/JVI.02865-15

Guseva, S., Milles, S., Blackledge, M., \& Ruigrok, R. (2019, August 21). The nucleoprotein and phosphoprotein of measles virus. Frontiers in Microbiology., 10(1832), 1-10. doi:10.3389/fmicb.2019.01832

Gutsche, I., Desfosses, A., Effantin, G., Ling, W., Haupt, M., Ruigrok, R., Schoehn, G. (2015, May 8). Near-atomic cryo-EM structure of the helical measles virus nucleocapsid. . Science., 348(6235), 704-709. doi:10.1126/science.aaa5137

Hahné, S., Nic Lochlainn, L., van Burgel, N., Kerkhof, J., Sane, J., Bing Yap, K., \& van Binnendijk, R. (2016, November 15). Measles outbreak among previously immunized healthcare workers, the Netherlands, 2014. . The Journal of Infectious Diseases., 1-7. doi:10.1093/infdis/jiw480

Hashiguchi, T., Maenaka, K., \& Yanagi, Y. (2011, December 11). Measles virus hemagglutinin: structural insights into cell entry and measles vaccine. Frontiers in Microbiology. , 2(247), 1-7. doi:10.3389/fmicb.2011.00247

Hashiguchi, T., Ose, T., Kubota, M., Maita, N., Kamishikiryo, J., Maenaka, K., \& Yanagi, Y. (2011, February). Structure of the measles virus hemagglutinin bound to its cellular receptor SLAM. Nature Structural \& Molecular Biology., 18(2), 135-133. doi:10.1038/nsmb.1969

Huang, B., Gomez-Rodriguez, J., Preite, S., Garrett, L., Harper, U., \& Schwartzberg, P. (2016). CRISPR-mediated triple knockout of SLAMF1, SLAMF5 and SLAMF6 supports positive signaling roles in NKT cell development. PLoS ONE, 11(6), 1-19. doi:10.1371/journal.pone.0156072

Ito, M., Iwasaki, M., Takeda, M., Nakamura, T., Yanagi, Y., \& Ohno, S. (2013, September). Measles virus nonstructural C protein modulates viral RNA polymerase activity by interacting with host protein SHCBP1. Journal of Virology., 87(17), 9633-9642. doi:10.1128/JVI.00714-13

Johansson, K., Bourhis, J.-M., Campanacci, V., Cambillau, C., Canard, B., \& Longhi, S. (2003, August 21). Crystal structure of the measles virus phosphoprotein domain responsible for the induced folding of the C-terminal domain of the nucleoprotein. The Journal of Biological Chemistry., 278(45), 44567-44573. doi: 10.1074/jbc.M308745200

Kimura, H., Saitoh, M., Kobayashi, M., Ishii, H., Saraya, T., Kurai, D., Takeda, M. (2015, July 01). Molecular evolution of haemagglutinin (H) gene in measles virus. Nature Scientific Reports. , 5(11648), 1-10. doi:10.1038/srep11648

Kong, L., Sun, L., Zhang, H., Liu, Q., Qin, L., Shi, G., . . . Ge, B.-X. (2009, August 20). An essential role for RIG-I in Toll-like receptor-stimulated phagocytosis. Cell Host \& Microbe. , 6, 150-161. doi:10.1016/j.chom.2009.06.008

Lai, W.-S., Lin, Y.-Y., Wang, C.-H., \& Chen, H.-C. (2017). Measles: a missed cause of acute tonsilitis. Ear, Nose \& Throat Journal., 96(10-11), E55. 

doi:10.1177/0145561317096010-1111

Liljeeros, L., Huiskonen, J., Ora, A., Susi, P., \& Butcher, S. (2011, November 1). Electron cryotomography of measles virus reveals how matrix protein coats the ribonucleocapsid within intact virions. PNAS, 108(44), 18085-18090. doi:10.1073/pnas.1105770108/-/DCSupplemental

Lin, L.-T., \& Richardson, C. (2016, September 20). The host cell receptors for measles virus and their interaction with the viral hemagglutinin (H) protein. Viruses, 8(250), 1-29. doi:10.3390/v8090250

Liszewski, M., \& Atkinson, J. (2015). Complement regulator CD46: genetic variants and disease associations. Human Genomics, 9(7), 1-13. doi:10.1186/s40246-015-0029-z

Manchester, M., Smith, K., Eto, D., Perkin, H., \& Torbett, B. (2002, July). Targeting and hematopoietic suppression of human CD34+ cells by measles virus. Journal of Virology., 76(13), 6636-6642. doi:10.1128/JVI.76.13.6636–6642.2002

Montella, S., Santamaria, F., Maglione, M., \& Ciofi degli Atti, M. (2009, February). [Measles and its secondary pulmonary complications: prevention is better than treatment]. Ann. Ig., 21(1), 17-27. Retrieved March 31, 2020, from 

https://www.ncbi.nlm.nih.gov/pubmed/19385330

Moss, W. (2017). Measles. The Lancet., 390(10111), 2490-2502. doi:10.1016/s0140-6736(17)31463-0

Mura, M., Combredet, C., Najburg, V., Sanchez David, R., Tangy, F., \& Komarova, A. (2017, August 2). Nonencapsidated 5' copy-back defective interfering genomes produced by recombinant measles viruses are recognized by RIG-I and LPGP2 but not MDA5. Journal of Virology. , 91(20), 1-22. doi:10.1128/JVI.00643-17

Nakashima, M., Shirogane, Y., Hashiguchi, T., \& Yanagi, Y. (2013, March 22). Mutations in the putative dimer-dimer interfaces of the measles virus hemagglutinin head domain affect membrane fusion triggering. The Journal of Biological Chemistry, 288(12), 8085-8091. doi:10.1074/jbc.M112.427609

Neppert, B. (1994). Masern-Retinitis bei einem immunkompetenten Kind. Klin. Monatsbl. Augenheilkd., 205, 156-160. doi:10.1055/s-2008-1045509

NIH. (2020, March 22). AFDN afadin, adherens junction formation factor [ Homo sapiens (human) ]. Retrieved March 29, 2020, from NCBI: Gene:

https://www.ncbi.nlm.nih.gov/gene/4301

NIH. (2020, March 13). CD46 CD46 molecule [ Homo sapiens (human) ]. Retrieved March 19, 2020, from NCBI: Gene: https://www.ncbi.nlm.nih.gov/gene/4179

NIH. (2020, March 29). DDX58 DExD/H-box helicase 58 [ Homo sapiens (human) ]. Retrieved March 30, 2020, from NCBI: Gene: https://www.ncbi.nlm.nih.gov/gene/23586

NIH. (2020, March 22). DHX58 DExH-box helicase 58 [ Homo sapiens (human) ]. Retrieved March 31, 2020, from NCBI: Gene: https://www.ncbi.nlm.nih.gov/gene/79132

NIH. (2020, March 29). IFIH1 interferon induced with helicase C domain 1 [ Homo sapiens (human) ]. Retrieved March 30, 2020, from NCBI: Gene:

https://www.ncbi.nlm.nih.gov/gene/64135

NIH. (2020, March 13). MAVS mitochondrial antiviral signaling protein [ Homo sapiens (human) ]. Retrieved March 30, 2020, from NCBI: Gene:

https://www.ncbi.nlm.nih.gov/gene/57506

NIH. (2020, March 13). NECTIN4 nectin cell adhesion molecule 4 [ Homo sapiens (human) ]. Retrieved March 29, 2020, from NCBI: Gene:

https://www.ncbi.nlm.nih.gov/gene/81607

NIH. (2020, March 13). SLAMF1 signaling lymphocytic activation molecule family member 1 [ Homo sapiens (human) ]. Retrieved March 18, 2020, from NCBI: Gene: https://www.ncbi.nlm.nih.gov/gene/6504

Nishiwada, S., Sho, M., Yasuda, S., Shimada, K., Yamato, I., Akahori, T., Nakajima, Y. (2015). Nectin-4 expression contributes to tumor proliferation, angiogenesis and patient prognosis in human pancreatic cancer. Journal of Experimental \& Clinical Cancer Research., 34(30), 1-9. doi:10.1007/s00018-014-1763-4

Obam Mekanda, F.-M., Gwladys Monamele, C., Simo Nemg, F., Yonga, G., Ouapi, D., Penlap Beng, V., Demanou, M. (2019, September 25). Molecular characterization of measles virus strains circulating in Cameroon during the 2013-2016 epidemics. . PLoS ONE, 14(9), 1-10. doi:10.1371/journal.pone.0222428

Okada, H., Kobune, F., Sato, T., Kohama, T., Takeuchi, Y., Abe, T., Tashiro, M. (1999, December 18). Extensive lymphopenia due to apoptosis of uninfected lymphocytes in acute measles patients. Archives of Virology., 145, 905-920. Retrieved March 31, 2020

Orren, A., Kipps, A., Moodie, J., Beatty, D., Dowdle, E., \& McIntyre, J. (1981, June). Increased susceptibility fo herpes simplex virus infections in children with acute measles. Infection and Immunity., 31(1), 1-6. Retrieved March 31, 2020, from
https://www.ncbi.nlm.nih.gov/pmc/articles/PMC351743/pdf/iai00165-0021.pdf

Parks, C., Witko, S., Kotash, C., Lin, S., Sidhu, M., \& Udem, S. (2006, January 25). Role of V protein RNA binding in inhibition of measles virus minigenome replication. Virology. , 348, 96-106. doi:10.1016/j.virol.2005.12.018

Paul Duprex, W., Collins, F., \& Kima, B. (2002, July). Modulating the function of the measles virus RNA-dependet RNA polymerase by insertion of green fluorescent protein into the open reading frame. Journal of Virology. , 76(14), 7322-7328. doi:10.1128/JVI.76.14.7322–7328.2002

Pfaller, C., \& Conzelmann, K.-K. (2008, December). Measles virus V protein is a decoy substrate for IKB kinase alpha and prevents toll-like receptor 7/9-mediated interferon induction. Journal of Virology., 82(24), 12365-12373. doi:10.1128/JVI.01321-08

Plattet, P., Alves, L., Herren, M., \& Aguilar, H. (2016, April 21). Measles virus fusion protein: structure, function and inhibition. . Viruses. , 8(112), 1-30. doi:10.3390/v8040112 

Poeck, H., Bscheider, M., Gross, O., Finger, K., Roth, S., Rebsamen, M., Ruland, J. (2010, January). Recognition of RNA virus by RIG-I results in activation of CARD9 and inflammasome signaling for interleukin 1B production. Nature Immunology., 11(1), 63-71. doi:10.1038/ni.1824

Pons, T., Gómez, R., Chinea, G., \& Valencia, A. (2003). Beta propellers: associated functions and their role in human diseases. Current Medicinal Chemistry, 10, 505-524. doi:10.2174/0929867033368204

PubChem. (2020). 4-Methyl-5-hydroxyethylthiazole phosphate. Retrieved March 30, 2020, from NIH: https://pubchem.ncbi.nlm.nih.gov/compound/1137

Rafat, C., Klouche, K., Ricard, J.-D., Messika, J., Roch, A., Machado, S., Gaudry, S. (2013, September). Severe measles infection. The spectrum of disease in 36 clinicaly ill adult patients. Medicine., 92(5), 257-273. doi:10.1097/MD.0b013e3182a713c2

Rawlings, J., Rosler, K., \& Harrison, D. (2004). The JAK/STAT signaling pathway. Journal of Cell Science, 117, 1281-1283. doi:10.1242/jcs.00963

Retief, F., \& Cilliers, L. (2010, April). Measles in antiquity and the Middle Ages. SAMJ, 100(4), 216-217. doi:10.7196/samj.3504

Rota, P., Moss, W., Takeda, M., de Swart, R., Thompson, K., \& Goodson, J. (2016, July 14). Measles. Nature Reviews | Disease Primers, 2, 1-16. doi:10.1038/nrdp.2016.49

Royal Society of Chemistry. (2020). 3-[[5-Bromanyl-1-(3-Methylsulfonylpropyl)benzimidazol-2-Yl]methyl]-1-Cyclopropyl-Imidazo[4,5-C]pyridin-2-One. Retrieved March 19, 2020, from ChemSpider: 

https://www.chemspider.com/Chemical-Structure.59052151.html?rid=58fe16b5-deb7-4998-8044-7ab27350bd9c

Royal Society of Chemistry. (2020). 4-Nitro-2-[(phenylacetyl)amino]benzamide. Retrieved March 19, 2020, from ChemSpider: http://www.chemspider.com/Chemical-Structure.23274280.html
Royal Society of Chemistry. (2020). JNJ-2408068. Retrieved March 19, 2020, from ChemSpider: 

http://www.chemspider.com/Chemical-Structure.445986.html

Rui, L., Drennan, A., Ceribelli, M., Zhi, F., Wright, G., Huang, D., Staudt, L. (2016, October 31). Epigenetic gene regulation by Janus kinase 1 in diffuse large B-cell lymphoma. Proceedings of the National Academy of Sciences of the United States of America., 113(46), E7260-E7267. doi:10.1073/pnas.1610970113

Saitoh, M., Takeda, M., Gotoh, K., Takeuchi, F., Sekizuka, T., Kuroda, M.,  Kimura, H. (2012, November 29). Molecular evolution of hemagglutinin (H) gene in measles virus genotypes D3, D5, D9, and H1. . PLoS ONE, 7(11), 1-6. 

doi:10.1371/journal.pone.0050660

Samanta, D., \& Almo, S. (2015). Nectin family of cell-adhesion molecules: structural and molecular aspects of function and specificity. Cellular and Molecular Life Sciences., 72, 645-658. doi:10.1007/s00018-014-1763-4

Sanchez David, R., Combredet, C., Najburg, V., Millot, G., Beauclair, G., Schwikowski, B., Komarova, A. (2019). LGP2 binds to PACT to regulate RIG-I-and MDA5-mediated antiviral responses. Science Signaling., 12(eaar3993), 1-14. 

doi:10.1126/scisignal.aar3993

Santiago, C., Celma, M., Stehle, T., \& Casasnovas, J. (2009, December 13). Structure of the measles virus hemagglutinin bound to the CD46 receptor. . Nature Structural \& Molecular Biology. , 17(1), 124-131. doi:10.1038/nsmb.1726

Strebel, P., \& Orenstein, W. (2019). Measles. New England Journal of Medicine. doi:10.1056/nejmcp1905181

Tahara, M., Bürckert, J.-P., Kanou, K., Maenaka, K., Muller, C., \& Takeda, M. (2016, August 02). Measles virus hemagglutinin protein epitopes: the basis of antigenic stability. Viruses. , 8(216), 1-15. doi:10.3390/v8080216 

Tait Wojno, E., Hunter, C., \& Stumhofer, J. (2019, April 16). The immunobiology of the Interleukin-12 family: room for discovery. Immunity, 50, 851-871. doi:10.1016/j.immuni.2019.03.011

Takeuchi, K., Miyajima, N., Nagata, N., Takeda, M., \& Tashiro, M. (2003). Wild-type measles virus induces large syncytium formation in primary human small airway epithelial cells by a SLAM(CD150)-independent mechanism. Virus Research., 94, 11-16. doi:10.1016/S0168-1702(03)00117-5

Terada, T., \& Yokoyama, S. (2015). Chapter Thirteen - Escherichia coli cell-free protein synthesis and isotope labeling of mammalian proteins. In Methods in Enzymology. (Vol. 565, pp. 311-345). Yokohama, Japan: Elsevier Inc. doi:10.1016/bs.mie.2015.08.035

Thomas, G. (2002, October). Furin at the cutting edge: from protein traffic to embryogenesis and disease. Nature Reviews | Molecular Cell Biology. , 3, 753-767. doi:10.1038/nrm934

Tisoncik, J., Korth, M., Simmons, C., Farrar, J., Martin, T., \& Katze, M. (2012). Into the eye of the citokine storm. Microbiology and Molecular Biology Reviews., 16-32. doi:10.1128/MMBR.05015-11

TyersLab. (2020). FURIN. Retrieved March 19, 2020, from BioGRID: Result Summary: https://thebiogrid.org/111082/summary/homo-sapiens/furin.html

Ul Haq Lodhi, O., Imam, S., Umer, M., \& Zafar, R. (2017, August 30). A rare coincidence of measles with typhoid fever. Cureus., 9(8), 1-7. doi:10.7759/cureus.1630

Wang, J., Wu, S., Jin, X., Li, M., Chen, S., Teeling, J., Gu, J. (2008). Retinoic acid-inducible gene-I mediates late phase induction of TNFA by lipopolysaccharide. . The Journal of Immunology. , 180, 8011-8019. doi:10.4049/jimmunol.180.12.8011

Wang, L.-F., Collins, P., Fouchier, R., Kurath, G., Lamb, R., Randall, R., \& Rima, B. (2018, July). Paramyxoviridae: Taxonomy - Then and Now. Retrieved March 16, 2020, from ICTV 9th Report (2011).:

https://talk.ictvonline.org/ictv-reports/ictv\_9th\_report/negative-sense-rna-viruses-2011/w/negrna\_viruses/199/paramyxoviridae

Weisstein, E. (2020). Gershgorin Circle Theorem. Retrieved April 02, 2020, from MathWorld--A Wolfram Web Resource:

https://mathworld.wolfram.com/GershgorinCircleTheorem.html

Whalen, T., Klos, J., Kovalcik, P., \& Cross, G. (1980, June). Measles and appendicitis. Am. Surg., 46(7), 412-3. doi:7447177

Wright, P., \& Dyson, H. (2015, January). Intrinsically disordered proteins in cellular signaling and regulation. Nature Reviews, 16, 18-30. doi:10.1038/nrm3920

Xie, S., Bahl, K., Reinecke, J., Hammond, G., Naslavsky, N., \& Caplan, S. (2015, October 28). The endocytic recycling compartment maintains cargo segregation acquired upon exit from the sorting endosome. Molecular Biology of the Cell, 27, 108-127. doi:10.1091/mbc.E15-07-0514

Yamamoto, H., Fara, A., Dasgupta, P., \& Kemper, C. (2013). CD46: The 'multitasker' of complement proteins. The International Journmal of Biochemistry \& Cell Biology. , 45, 2808-2820. doi:10.1016/j.biocel.2013.09.016

Yurchenko, M., Skjesol, A., Ryan, L., Mary Richard, G., Kumaran Kandasamy, R., Wang, N., Espevik, T. (2018). SLAMF1 is required for TLR4-mediated TRAM-TRIF-dependent signaling in human macrophages. J. Cell. Biol. , 1-19. doi:10.1083/jcb.201707027

Zeng, W., Sun, L., Jiang, X., Chen, X., Hou, F., ADhikari, A., Chen, Z. (2010, April 16). Reconstitution of the RIG-I pathway reveals a pivotal role of unanchored polyubiquitin chains in innate immunity. Cell., 141(2), 315-330. doi:10.1016/j.cell.2010.03.029.

Zhou, L., Sun, S., Xu, L., Yu, Y., Zhang, T., \& Wang, M. (2019). DExH-Box helicase 58 enhances osteoblast differentuation of osteoblastic cells via Wnt/beta-catenin signaling. Biochemical and Biophysical Research Communications., 511, 307-311. doi:10.1016/j.bbrc.2019.02.039

Zill. (2009). Ecuaciones diferenciales con aplicaciones de modelado. Ciudad de México: Cengage Learning.

\end{document}